\documentclass[11pt,a4paper]{article}
\usepackage[margin=25mm]{geometry}
\usepackage{iftex}
\ifPDFTeX
  \usepackage[T1]{fontenc}
  \usepackage[utf8]{inputenc}
  \usepackage{lmodern}
  \newcommand{\coverfont}{\fontfamily{ptm}\selectfont}
\else
  \usepackage{fontspec}
  \newfontfamily\coverfont{texgyretermes-regular.otf}[
    BoldFont=texgyretermes-bold.otf,
    ItalicFont=texgyretermes-italic.otf,
    BoldItalicFont=texgyretermes-bolditalic.otf]
\fi
\usepackage{amsmath,amssymb,textcomp}
\usepackage{microtype}
\usepackage{graphicx}
\usepackage[table]{xcolor}
\usepackage{array,longtable,booktabs,calc}
\usepackage[framemethod=default]{mdframed}
\usepackage{pdflscape}
\usepackage{eso-pic}
\usepackage{needspace}
\usepackage{enumitem}
\usepackage{caption}
\usepackage{xurl}
\usepackage[unicode,pdfencoding=auto,psdextra]{hyperref}
\usepackage{bookmark}
\usepackage{footnotehyper}
\hypersetup{
  colorlinks=true,linkcolor=black,citecolor=black,urlcolor=blue!45!black,
  pdftitle={Transparency Disclosures for Chain-of-Thought Monitorability},
  pdfauthor={Jacob Davies and Theo Bearman}
}
\definecolor{TableHeading}{HTML}{936348}
\definecolor{SelectedRow}{HTML}{F2E8DD}
\definecolor{BodyRow}{HTML}{F3F3F3}
\definecolor{TableGrid}{HTML}{808080}
\definecolor{MainTextLinkColor}{HTML}{005EA8}
\newlength{\tableusablewidth}
\setlist{itemsep=4pt,topsep=5pt}

\makeatletter
\newenvironment{indicatorpanel}{%
  \begin{savenotes}%
  \let\mdf@footnoteinput\relax
  \let\mdf@footnoteoutput\relax
  \begin{mdframed}[
    nobreak=true,
    backgroundcolor=black!5,
    linecolor=black!50,
    linewidth=0.5pt,
    roundcorner=0pt,
    leftmargin=0pt,rightmargin=0pt,
    innerleftmargin=10pt,innerrightmargin=10pt,
    innertopmargin=8pt,innerbottommargin=8pt,
    skipabove=8pt,skipbelow=8pt]
  \small\setlength{\parskip}{6pt}%
}{%
  \end{mdframed}%
  \end{savenotes}%
}
\makeatother
\newif\ifShowPageNumber
\newif\ifLandscapePage
\newcommand{\PageNumberText}{\normalfont\normalsize\color{black}\thepage}
\newsavebox{\PageNumberBox}
\AddToShipoutPictureFG{%
  \ifShowPageNumber
    \begingroup
    \setlength{\unitlength}{1pt}%
    \sbox{\PageNumberBox}{\PageNumberText}%
    \AtPageLowerLeft{%
      \ifLandscapePage
        \put(\LenToUnit{\dimexpr\paperwidth-15mm-\ht\PageNumberBox\relax},%
             \LenToUnit{\dimexpr\paperheight-25mm\relax})%
          {\rotatebox{90}{\makebox[0pt][r]{\usebox{\PageNumberBox}}}}%
      \else
        \put(\LenToUnit{\dimexpr\paperwidth-25mm\relax},%
             \LenToUnit{15mm})%
          {\makebox[0pt][r]{\usebox{\PageNumberBox}}}%
      \fi
    }%
    \endgroup
  \fi
}
\begin{document}
\hypersetup{pageanchor=false}
\newgeometry{top=17mm,bottom=22mm,left=17mm,right=17mm}
\begin{titlepage}
  \coverfont
  \setlength{\parskip}{0pt}
  \setlength{\parindent}{0pt}
  \hrule height 3pt
  \vspace{11pt}
  {\centering\fontsize{17}{20}\selectfont\bfseries
    Transparency Disclosures for\\
    Chain-of-Thought Monitorability\par}
  \vspace{11pt}
  \hrule height 0.4pt
  \vspace{11pt}
  {\centering\fontsize{11}{14}\selectfont\bfseries
    Jacob Davies\hyperlink{cover-author-note}{\textsuperscript{*}}\qquad
    Theo Bearman\textsuperscript{1}\par}
  \vspace{7pt}
  {\centering\fontsize{11}{13}\selectfont
    \textsuperscript{1}Institute for AI Policy and Strategy\par}
  \vspace{35pt}
  {\centering\fontsize{12}{14}\selectfont\bfseries Abstract\par}
  \vspace{12pt}
  \noindent\hspace*{0.09\textwidth}\begin{minipage}{0.82\textwidth}
    \fontsize{11}{12.8}\selectfont
    Chain-of-thought (CoT) monitoring is one of the most valuable tools for monitoring advanced AI systems, but it is under threat. We argue that frontier AI developers should report specific indicators to make these threats trackable and to better facilitate coordination to preserve CoT monitorability. We identify five indicators that are both highly relevant and reasonably feasible to disclose, one for each link in a five-step chain on which CoT monitoring depends. The first, third-party CoT access and base architecture specifications, also underpins external verification of the rest. The others span CoT quality (naturalness and faithfulness) and monitoring effectiveness (detection and uplift). An audit of current disclosures by Anthropic, OpenAI, and Google DeepMind shows that no developer covers the full set, but that every indicator has been reported by at least one frontier developer. Because current disclosures rely largely on developers’ own evaluations, we also recommend that a cross-developer audit be repeated with deeper access than public disclosures allow, either by embedded evaluators working within developers or by an Independent Verification Organization.
  \end{minipage}\par
  \vfill
  \noindent\rule{0.25\textwidth}{0.3pt}\par
  \vspace{3pt}
  {\fontsize{9}{11}\selectfont
    \hypertarget{cover-author-note}{\textsuperscript{*}}Jacob Davies conducted this work while
    participating in the ERA Fellowship, Cambridge, UK\par}
\end{titlepage}
\restoregeometry
\pagenumbering{roman}
\global\ShowPageNumbertrue
\hypersetup{pageanchor=true}
\tableofcontents
\clearpage
\pagenumbering{arabic}
\Needspace{6\baselineskip}

\section{Editorial Note}\label{editorial-note}

This report was first circulated as a preprint paper in July 2026 that was accepted to the technical AI governance workshop at the International Conference on Machine Learning (ICML) 2026.\footnote{Davies and Guest, ``\href{https://openreview.net/forum?id=H5FSdlrdZs}{Transparency Disclosures for Chain-of-Thought Monitorability}.''} Oliver Guest, one of the original co-authors, undertook this work while at IAPS. The analysis and cross-developer audits in that version were completed based on the most recent model or system cards published as of May 2026.

Since the July 2026 version of this report, the field of chain-of-thought (CoT) monitorability has seen significant developments, most notably OpenAI's disclosure that GPT-6 Astra ``shows a substantial decrease in chain-of-thought monitorability compared to previous models.''\footnote{OpenAI, ``\href{https://deploymentsafety.openai.com/gpt-6-astra}{GPT-6 Astra System Card}.''} This version has been updated to reflect this disclosure and related developments through October 2026.

This version primarily adds evidence supporting the report's arguments rather than changing them. Sections 2 and 4 in the original report have been combined to form the new {\begingroup\hypersetup{linkcolor=MainTextLinkColor,urlcolor=MainTextLinkColor}\hyperref[a-framework-for-chain-of-thought-monitoring]{Section 2}\endgroup{}} in this version, and three sections have been substantively updated: {\begingroup\hypersetup{linkcolor=MainTextLinkColor,urlcolor=MainTextLinkColor}\hyperref[the-cot-must-exist-and-be-necessary-for-complex-reasoning]{Section 2 (The CoT Must Exist and Be Necessary For Complex Reasoning)}\endgroup{}} now accounts for the possibility that a model's CoT may not fully capture its reasoning mechanisms; {\begingroup\hypersetup{linkcolor=MainTextLinkColor,urlcolor=MainTextLinkColor}\hyperref[the-cot-must-be-in-readable-human-language]{Section 2 (The CoT Must be in Readable Human Language)}\endgroup{}} now specifies a rubric-based legibility score in place of the proportion of reasoning tokens classified as human language; and the cross-developer audit in {\begingroup\hypersetup{linkcolor=MainTextLinkColor,urlcolor=MainTextLinkColor}\hyperref[cross-developer-audit-of-current-disclosures]{Section 4}\endgroup{}} reflects disclosures through mid-October 2026 and includes a recommendation that the audit be repeated with deeper access by embedded evaluators or an Independent Verification Organization. Otherwise, the framework, conditions and indicators, and the disclosures scoring methodology ({\begingroup\hypersetup{linkcolor=MainTextLinkColor,urlcolor=MainTextLinkColor}\hyperref[method-for-identifying-monitorability-indicators]{Section 3}\endgroup{}}) and results ({\begingroup\hypersetup{linkcolor=MainTextLinkColor,urlcolor=MainTextLinkColor}\hyperref[full-scored-framework]{Appendix (Full Scored Framework)}\endgroup{}}) are substantively unchanged.

\clearpage

\Needspace{6\baselineskip}

\section{Executive Summary}\label{executive-summary}

Chain-of-thought (CoT) reasoning provides a window into the decision-making of frontier AI systems. Monitoring this decision-making is one of the most effective oversight tools for current systems, and one with the potential to scale to more advanced future systems.\footnote{OpenAI, ``\href{https://openai.com/index/chain-of-thought-monitoring/}{Detecting misbehavior in frontier reasoning models}''; Korbak et al., ``\href{https://arxiv.org/abs/2507.11473}{Chain of Thought Monitorability: A New and Fragile Opportunity for AI Safety}.''} However, there are threats to this window: it may become obscured as CoT becomes unfaithful,\footnote{Barez et al., ``\href{https://aigi.ox.ac.uk/publications/chain-of-thought-is-not-explainability/}{Chain-of-Thought Is Not Explainability}.''} encoded,\footnote{Baker et al., ``\href{https://arxiv.org/abs/2503.11926}{Monitoring Reasoning Models for Misbehavior and the Risks of Promoting Obfuscation}''; Skaf et al., ``\href{https://arxiv.org/abs/2506.01926}{Large language models can learn and generalize steganographic chain-of-thought under process supervision}.''} or illegible,\footnote{Schoen et al., ``\href{https://arxiv.org/abs/2509.15541}{Stress Testing Deliberative Alignment for Anti-Scheming Training}.''} and frontier models can increasingly reason without CoT---or at least rely on it less---without degrading capability. The window may narrow if frontier models increasingly rely on less transparent reasoning mechanisms or disappear entirely if they stop using CoT altogether.\footnote{See for example: Hao et al., ``\href{https://arxiv.org/abs/2412.06769}{Training Large Language Models to Reason in a Continuous Latent Space}''; Lu et al., ``\href{https://openreview.net/forum?id=roIQdXMuEj}{Latent Chain-of-Thought? Decoding the Depth-Recurrent Transformer}''; Geiping et al., ``\href{https://arxiv.org/abs/2502.05171}{Scaling up Test-Time Compute with Latent Reasoning: A Recurrent Depth Approach}''; Wang et, al., ``\href{https://arxiv.org/abs/2608.08888}{Full-bandwidth transformer}.''}

The OpenAI/Hugging Face incident, arguably the most significant publicly acknowledged case of unsanctioned AI agent activity to date, illustrates the stakes of losing the insights CoT monitoring provides. OpenAI has said its CoT monitoring system would likely have caught the agent activity had it been running at the time, but that is only an effective safeguard as long as a model's CoT remains monitorable.\footnote{OpenAI, ``\href{https://openai.com/index/hugging-face-incident-and-the-road-ahead/}{The Hugging Face incident and the road ahead}''; METR, ``\href{https://metr.org/blog/2026-08-26-openai-hugging-face-incident-investigation/}{Brief independent investigation of agents' behavior, reasoning and collaboration in the OpenAI / Hugging Face hacking incident}.''} Future incidents of similar or greater impact could go un-detected if models gain the ability to evade CoT monitoring. Moreover, the challenges of monitoring can be exacerbated by training practices that penalize specific content in the CoT, creating optimization pressure for models to hide their bad intentions rather than change the underlying behavior.\footnote{Baker et al., ``\href{https://arxiv.org/abs/2503.11926}{Monitoring Reasoning Models for Misbehavior and the Risks of Promoting Obfuscation}''; Korbak et al., ``\href{https://arxiv.org/abs/2507.11473}{Chain of Thought Monitorability: A New and Fragile Opportunity for AI Safety}.''}

Recent evidence suggests that the ability to understand and evaluate models' reasoning based on CoT is weakening in frontier AI systems. For example, in September 2026, OpenAI released a new model, GPT-6 Astra, that showed ``a substantial decrease in chain-of-thought monitorability compared to previous models,'' and the set of tasks it can complete without any CoT ``is greatly expanded''.\footnote{OpenAI, ``\href{https://deploymentsafety.openai.com/gpt-6-astra}{GPT-6 Astra System Card}.''} Anthropic has raised similar concerns, particularly regarding its most powerful Mythos-class models.\footnote{Anthropic, ``\href{https://www-cdn.anthropic.com/d00db56fa754a1b115b6dd7cb2e3c342ee809620.pdf}{System Card: Claude Fable 5 \& Claude Mythos 5}''; Anthropic, ``\href{https://www-cdn.anthropic.com/0339e6a7c5c7b87f5c07798616dc32c215d14235/Claude\%20Fable\%205.1\%20&\%20Claude\%20Mythos\%205.1\%20System\%20Card.pdf}{System Card: Claude Fable 5.1 \& Claude Mythos 5.1}.''} Google DeepMind has said the monitorability of their newest frontier AI model, Gemini 4 Argon, is ``more challenging compared to past models,'' though the CoT remains ``legible, informative, and free of spontaneous obfuscation.''\footnote{Google DeepMind, ``\href{https://storage.googleapis.com/deepmind-media/gemini/gemini_4_safety_report.pdf}{Gemini 4 Argon Safety \& Security Report}.''} Media reporting also suggested that OpenAI's GPT-6 Astra leveraged a ``looped transformer'' architecture\footnote{Kahn, J. ``\href{https://fortune.com/2026/09/03/reports-openais-astra-model-uses-a-new-more-efficient-ai-architecture-alarms-ai-safety-experts-who-worry-the-method-makes-models-harder-to-control/}{Safety experts warn novel design of OpenAI's Astra model could make future AI agents harder to monitor}.''} that, while retaining CoT, lets the model do more reasoning without needing to use it,\footnote{Geiping et al., ``\href{https://arxiv.org/abs/2502.05171}{Scaling up Test-Time Compute with Latent Reasoning: A Recurrent Depth Approach}.''} though OpenAI has not categorically confirmed this.

This report argues that frontier AI developers should regularly disclose a small set of specific indicators so the trajectory of CoT monitorability can be tracked by independent evaluators and developers have stronger incentives to preserve it. We propose that CoT monitoring depends on a chain of five conditions and, for each, identify one indicator that is both highly relevant and reasonably feasible to disclose. The five indicators are:

\begin{enumerate}
\def\labelenumi{\arabic{enumi}.}
\item
  \textbf{Third-party CoT Access and Base Architecture Specifications.} Allowing access to a model's raw CoT to qualified third parties provides an initial warning sign about whether models continue to use CoT, and underpins external verification of the other indicators in our framework. To provide assurances that the CoT is necessary for complex reasoning capabilities, access should be complemented with relevant details of the model's base architecture specifications that can be independently validated.
\item
  \textbf{Naturalness.} To use CoT to monitor for harmful actions, monitors must be able to extract semantic meaning from it. A legibility score, rated on a standard rubric of how easily a human can follow the reasoning, can indicate whether and how far a model's reasoning is drifting from human-readable language.
\item
  \textbf{Faithfulness.} Access to a human-readable CoT is only helpful if its contents match the computation the model actually performed. Scores on a faithfulness suite covering a variety of failure modes can show whether a model's CoT reflects its actual reasoning.
\item
  \textbf{Detection.} CoT supports model oversight when a monitoring system detects dangerous behavior expressed through it. The proportion of harmful behaviors that evade the monitoring system indicates the extent to which a model might be able to take unsanctioned actions before it is stopped.
\item
  \textbf{Uplift.} Efforts to secure and expand CoT monitorability are justified only if access to the CoT actually improves a monitor's performance compared with having access to the model's actions alone. Understanding the performance gap between monitors with these two different levels of access quantifies the value CoT access provides for safety monitoring.
\end{enumerate}

We selected these from a list of 21 candidate indicators identified in recent CoT monitorability research literature, developer disclosures, and other sources. We then scored these for informativeness, technology readiness, external verifiability, developer tractability, and implementation feasibility. We do not prescribe specific evaluations for each indicator to preserve space for state-of-the-art approaches to be used and to reduce the scope for gaming.

We audited current disclosure practices among the three leading frontier AI developers---OpenAI, Anthropic, and Google DeepMind---and found that no developer reports the full set of indicators. That said, at least one developer has reported each individual indicator, showing that cross-sector reporting of the full set is feasible.\footnote{See for example: OpenAI, ``\href{https://deploymentsafety.openai.com/gpt-5-6}{GPT-5.6 System Card}''; OpenAI, ``\href{https://deploymentsafety.openai.com/gpt-6-astra}{GPT-6 Astra System Card}''; Anthropic, ``\href{https://www-cdn.anthropic.com/0339e6a7c5c7b87f5c07798616dc32c215d14235/Claude\%20Fable\%205.1\%20&\%20Claude\%20Mythos\%205.1\%20System\%20Card.pdf}{System Card: Claude Fable 5.1 \& Claude Mythos 5.1};''; Anthropic, ``\href{https://www-cdn.anthropic.com/870c8f525702625d2c62fc6dd04c857e3250bec1/Claude\%20Sonnet\%205.5\%20System\%20Card.pdf}{System Card: Claude Sonnet 5.5}''; Google DeepMind, ``\href{https://storage.googleapis.com/deepmind-media/gemini/gemini_3_pro_fsf_report.pdf}{Gemini 3 Pro Frontier Safety Framework Report}''; Google DeepMind, ``\href{https://storage.googleapis.com/deepmind-media/gemini/gemini_4_safety_report.pdf}{Gemini 4 Argon Safety \& Security Report}.''} Recent disclosures by OpenAI, Anthropic, and Google DeepMind that highlight evidence of increased challenges around the monitorability in their frontier models show that voluntary disclosures on these issues can happen.\footnote{OpenAI, ``\href{https://deploymentsafety.openai.com/gpt-6-astra}{GPT-6 Astra System Card}''; Anthropic, ``\href{https://www-cdn.anthropic.com/d00db56fa754a1b115b6dd7cb2e3c342ee809620.pdf}{System Card: Claude Fable 5 \& Claude Mythos 5}''; Anthropic, ``\href{https://www-cdn.anthropic.com/0339e6a7c5c7b87f5c07798616dc32c215d14235/Claude\%20Fable\%205.1\%20&\%20Claude\%20Mythos\%205.1\%20System\%20Card.pdf}{System Card: Claude Fable 5.1 \& Claude Mythos 5.1}''; Google DeepMind, ``\href{https://storage.googleapis.com/deepmind-media/gemini/gemini_4_safety_report.pdf}{Gemini 4 Argon Safety \& Security Report}.''} However, these disclosures are currently voluntary and based primarily on private evaluations and configurations, making it difficult for third parties to replicate the findings. In addition, a lack of standardized methods and evaluations for CoT monitorability makes independent cross-developer comparisons challenging. We recommend repeating this audit with deeper access than public disclosures allow, either by embedded evaluators working within developers or by an Independent Verification Organization.

We argue that frontier AI developers should disclose specific monitorability indicators as part of their existing system card and safety evaluation practices. Disclosure would make the trajectory of decreasing monitorability trackable before failures occur, and create reputational and coordination-related incentives to preserve monitorability.\footnote{Korbak et al., ``\href{https://arxiv.org/abs/2507.11473}{Chain of Thought Monitorability: A New and Fragile Opportunity for AI Safety}''; Delaney et al., ``\href{https://www.iaps.ai/research/policy-options-for-preserving-cot-monitorability}{Policy Options for Preserving Chain of Thought Monitorability}.''} Disclosing the CoT monitorability indicators we set out in this report can help policymakers and researchers track this tool's effectiveness in overseeing frontier AI systems, inform technical and policy responses when evidence shows it is weakening, and create reputational and coordination-related incentives to preserve it.\footnote{Korbak et al., ``\href{https://arxiv.org/abs/2507.11473}{Chain of Thought Monitorability: A New and Fragile Opportunity for AI Safety}''; Delaney et al., ``\href{https://www.iaps.ai/research/policy-options-for-preserving-cot-monitorability}{Policy Options for Preserving Chain of Thought Monitorability}.''}

\Needspace{6\baselineskip}

\section{1 \textbar{} The Value of Transparency About Monitorability Indicators}\label{the-value-of-transparency-about-monitorability-indicators}

Reliable data on indicators corresponding to chain-of-thought (CoT) monitorability would be useful for at least two reasons.

First, it would allow tracking of how monitorability is changing over time. Without disclosure, a shift toward obfuscated or latent reasoning, and thus reduced monitorability, could proceed gradually with no external signal until monitoring failures occur. Recent disclosures by OpenAI, Anthropic, and Google DeepMind that highlight evidence of increased challenges around the monitorability of their frontier models show that disclosures are possible.\footnote{OpenAI, ``\href{https://deploymentsafety.openai.com/gpt-6-astra}{GPT-6 Astra System Card}''; Anthropic, ``\href{https://www-cdn.anthropic.com/d00db56fa754a1b115b6dd7cb2e3c342ee809620.pdf}{System Card: Claude Fable 5 \& Claude Mythos 5}''; Anthropic, ``\href{https://www-cdn.anthropic.com/0339e6a7c5c7b87f5c07798616dc32c215d14235/Claude\%20Fable\%205.1\%20&\%20Claude\%20Mythos\%205.1\%20System\%20Card.pdf}{System Card: Claude Fable 5.1 \& Claude Mythos 5.1}''; Google DeepMind, ``\href{https://storage.googleapis.com/deepmind-media/gemini/gemini_4_safety_report.pdf}{Gemini 4 Argon Safety \& Security Report}.''} However, these disclosures are currently voluntary and based primarily on private evaluations and configurations, making it difficult for third parties to replicate the findings. In addition, a lack of standardized methods and evaluations for CoT monitorability makes independent cross-developer comparisons challenging. With indicator data, the trend would be much more visible and trackable. This would be useful for any policy or technical efforts that might be needed to mitigate risks posed by reduced CoT monitorability.

Second, disclosure would strengthen frontier developers' incentives to preserve CoT monitorability. Reporting the indicators would make it reputationally costly to let monitorability decline. Disclosure might also reduce a coordination problem: most frontier developers appear to regard monitorability as important,\footnote{Korbak et al., ``\href{https://arxiv.org/abs/2507.11473}{Chain of Thought Monitorability: A New and Fragile Opportunity for AI Safety}''; Frontier Model Forum, ``\href{https://www.frontiermodelforum.org/issue-briefs/chain-of-thought-monitorability/}{Chain of Thought Monitorability}.''} but each may hesitate to pay any costs associated with preserving monitorability without assurance that competitors will do the same.\footnote{We focus here on U.S. frontier developers.} Shared disclosure would provide such assurance. The case of GPT-6 Astra is particularly instructive. OpenAI's reporting of the model's degraded monitorability attracted significant scrutiny,\footnote{See for example: Van Hollen, ``Van Hollen presses OpenAI CEO Sam Altman on alarming new AI model claims, calls for risk assessment of AI capabilities,'' (retrieved from \href{https://web.archive.org/web/20261101015352/https://www.vanhollen.senate.gov/news/press-releases/van-hollen-presses-openai-ceo-sam-altman-on-alarming-new-ai-model-claims-calls-for-risk-assessment-of-ai-capabilities}{web archive}); Ford ``\href{https://www.transformernews.ai/p/openai-gpt-6-astra-might-be-too-powerful-to-understand-or-control}{OpenAI's GPT-6 Astra might be too powerful to understand or control}.''} demonstrating the reputational impact of reduced monitorability. However, OpenAI still released the model while acknowledging it had these characteristics, suggesting that reputational pressure alone may not be sufficient to preserve monitorability.\footnote{OpenAI, ``\href{https://deploymentsafety.openai.com/gpt-6-astra}{GPT-6 Astra System Card}.''}

Transparency disclosures are also a natural extension of existing reporting practices. Frontier AI developers already publish system cards,\footnote{See for example: Anthropic, ``\href{https://www-cdn.anthropic.com/0339e6a7c5c7b87f5c07798616dc32c215d14235/Claude\%20Fable\%205.1\%20&\%20Claude\%20Mythos\%205.1\%20System\%20Card.pdf}{System Card: Claude Fable 5.1 \& Claude Mythos 5.1}''; OpenAI, ``\href{https://deploymentsafety.openai.com/gpt-6-astra}{GPT-6 Astra System Card}''; Google DeepMind, ``\href{https://storage.googleapis.com/deepmind-media/Model-Cards/Gemini-3-Pro-Model-Card.pdf}{Gemini 3 Pro Model Card}.''} safety evaluations, and responsible scaling assessments that cover a range of risks from their AI systems. Adding monitorability indicators to these existing reporting frameworks would be relatively low in cost and would not constrain what developers do, only what they report. As such, the indicators described in this report would be comparatively easy for frontier developers to disclose, since they would not require significant workflow changes or the reporting of highly sensitive proprietary information beyond qualified third-parties.

\Needspace{6\baselineskip}

\section{\texorpdfstring{2 \textbar{} A Framework for Chain-of-Thought Monitoring }{2 \textbar{} A Framework for Chain-of-Thought Monitoring }}\label{a-framework-for-chain-of-thought-monitoring}

Our framework for chain-of-thought (CoT) monitorability requires five conditions to hold. Figure 1 presents these five conditions as a chain, with the failure mode that threatens each condition and the indicator that tracks whether the condition holds, each of which answers a distinct diagnostic question about the state of CoT monitorability. This section walks through the chain condition by condition, with each subsection ending in the proposed indicator for that link.

\begin{figure}[htbp]\centering

\includegraphics[width=\linewidth]{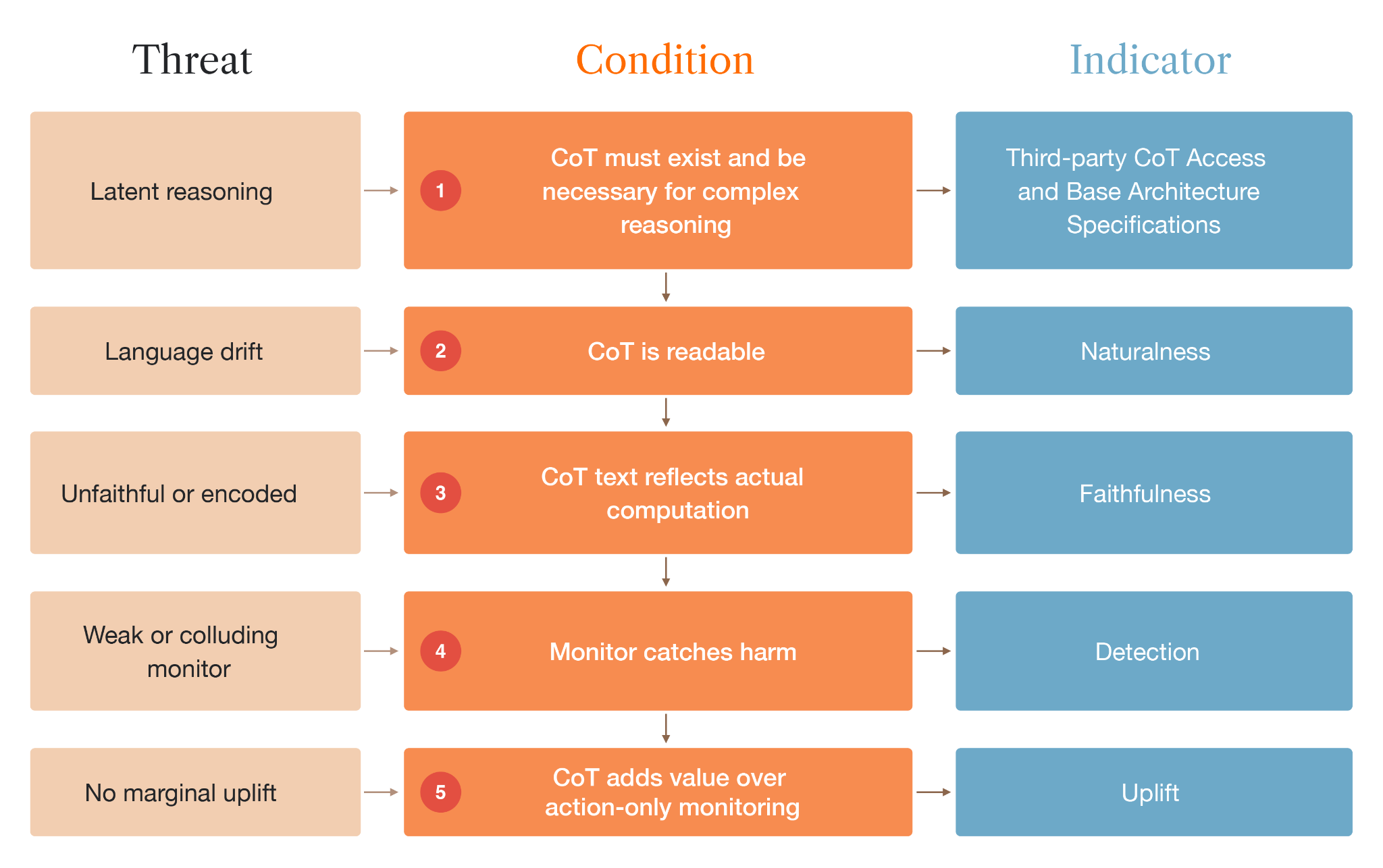}

\caption{CoT Monitoring as a Chain of Five Conditions}\label{fig:1}

\end{figure}

The framework specifies the information to be disclosed rather than the evaluation method, allowing developers to adopt the best available measurement at the time of each disclosure. We explain this design choice further in {\begingroup\hypersetup{linkcolor=MainTextLinkColor,urlcolor=MainTextLinkColor}\hyperref[discussion]{Section 5}\endgroup{}}.

\Needspace{12\baselineskip}

We do not claim the five conditions are exhaustive. However, every CoT failure mode we found in the literature breaks at least one of them (see Figure 2 and {\begingroup\hypersetup{linkcolor=MainTextLinkColor,urlcolor=MainTextLinkColor}\hyperref[threat-models-and-the-conditions-they-break]{Appendix, Threat Models and the Conditions They Break}\endgroup{}}), which gives us some confidence that the chain captures the principal ways monitoring can fail. Because the conditions form a chain, a single failure can register on more than one indicator. This is not a problem for the framework. When several indicators flag at once, the earliest broken link indicates where monitoring first failed; even when the precise cause cannot be isolated, knowing that monitorability has degraded is useful.

\begin{figure}[htbp]\centering

\begingroup
\small
\setlength{\tabcolsep}{4pt}
\arrayrulecolor{TableGrid}
\setlength{\arrayrulewidth}{0.5pt}
\setlength{\tableusablewidth}{\dimexpr\linewidth-4\tabcolsep-3\arrayrulewidth\relax}
\setlength{\LTpost}{0pt}
\renewcommand{\arraystretch}{1.24}
\begin{tabular}{|>{\raggedright\arraybackslash}p{0.32000\tableusablewidth}|
>{\raggedright\arraybackslash}p{0.68000\tableusablewidth}|}
\hline
\rowcolor{TableHeading}
\textcolor{white}{\bfseries Condition} & \textcolor{white}{\bfseries Representative Documented Failure Modes} \\ \hline
\rowcolor{BodyRow}
1. CoT Must Exist and Be Necessary for Complex Reasoning & Latent / continuous-state reasoning; architectural move off text CoT \\ \hline
\rowcolor{BodyRow}
2. CoT Is Readable & Language drift; neuralese compression \\ \hline
\rowcolor{BodyRow}
3. CoT Text Reflects Actual Computation & Post-hoc rationalisation; CoT as scaffolding; steganography; obfuscation under reinforcement learning; pathological CoT \\ \hline
\rowcolor{BodyRow}
4. Monitor Catches Harm & Weak monitor; monitor--model collusion; in-context evasion \\ \hline
\rowcolor{BodyRow}
5. CoT Adds Value Over Action-Only Monitoring & Shrinking CoT-vs-action gap; CoT too poor to help \\ \hline
\end{tabular}
\endgroup

\caption{Failure Modes Documented in the CoT Literature and the Condition Each One Breaks}\label{fig:failure-modes}

\vspace{4pt}\parbox{\linewidth}{\footnotesize \emph{Note: Each failure mode is associated with the earliest condition it breaks; some bear on more than one condition. Full table with sources in the {\begingroup\hypersetup{linkcolor=MainTextLinkColor,urlcolor=MainTextLinkColor}\hyperref[threat-models-and-the-conditions-they-break]{Appendix (Threat Models and the Conditions They Break)}\endgroup{}}.}}

\end{figure}

\Needspace{6\baselineskip}

\subsection{The CoT Must Exist and be Necessary for Complex Reasoning}\label{the-cot-must-exist-and-be-necessary-for-complex-reasoning}

Frontier reasoning models currently externalize their reasoning as text, and this externalization is what makes CoT monitoring possible. However, there are research models and architectures that demonstrate the feasibility of reasoning in continuous internal states, including those that bypass natural language text entirely.\footnote{See for example Hao et al., ``\href{https://arxiv.org/abs/2412.06769}{Training Large Language Models to Reason in a Continuous Latent Space}''; Lu et al., ``\href{https://openreview.net/forum?id=roIQdXMuEj}{Latent Chain-of-Thought? Decoding the Depth-Recurrent Transformer}''; Geiping et al., ``\href{https://arxiv.org/abs/2502.05171}{Scaling up Test-Time Compute with Latent Reasoning: A Recurrent Depth Approach}''; Wang et al., ``\href{https://arxiv.org/abs/2608.08888}{Full-bandwidth transformer}.''} Although no frontier model has been confirmed to deploy these ``latent reasoning'' architectures, there may be competitive incentives to remove the natural-language bottleneck, as it could yield significant efficiency and capability gains.\footnote{Delaney et al., ``\href{https://www.iaps.ai/research/policy-options-for-preserving-cot-monitorability}{Policy Options for Preserving Chain of Thought Monitorability}.''}

Some architectures rely on CoT for any reasoning beyond what a single forward pass of a model's neural network can achieve while others are geared toward completely removing the need for CoT as a reasoning mechanism, with a spectrum between the two. However, if CoT is undermined, regardless of the mechanism, it may not be clear whether it has occurred or what the impact is. Qualified third-party access to relevant model base architecture specifications would enable independent validation of whether models use reasoning mechanisms other than CoT, and whether the CoT remains necessary for complex reasoning capabilities.

Unless and until other methods can provide similar insight into a model's reasoning as CoT monitoring does, a shift to such latent architectures would collapse the rest of the chain outlined above. Its most immediate external signal would be the disappearance of inspectable CoT from appropriately permissioned developer application programming interfaces (APIs), which makes third-party access the practical indicator for this condition.

Access to the CoT would also be useful for verifying the other indicators. Because naturalness, faithfulness, detection, and uplift are all measured on the CoT, a third party cannot reproduce a developer's reported scores without the traces themselves. Third-party access to the CoT is therefore both the indicator for the first condition and the precondition for externally verifying the others.

There are legitimate reasons why frontier developers might be reluctant to make their model's raw CoTs and base architecture specifications accessible. For example, other developers could use this information to train their own models,\footnote{Anthropic, ``\href{https://www.anthropic.com/news/detecting-and-preventing-distillation-attacks}{Detecting and preventing distillation attacks}.''} and models sometimes surface dangerous information in the CoT that does not appear in the final output.\footnote{OpenAI, ``\href{https://openai.com/index/learning-to-reason-with-llms/}{Learning to reason with LLMs}.''} We therefore limit the recommendation to qualified third parties such as embedded evaluators, Independent Verification Organizations, governments, and accredited researchers as a way of balancing these considerations against the value of access.

There is already precedent for limited sharing of reasoning traces. For example, in the nonprofit AI evaluator METR's pilot frontier-risk assessment, participating developers shared their internal models, including raw chains of thought, with METR under agreed terms.\footnote{METR, ``\href{https://metr.org/blog/2026-05-19-frontier-risk-report/}{Frontier Risk Report (February to March 2026.)}''} Google DeepMind has also stated it may grant access to ``raw chain-of-thought traces to bound latent capabilities'' if it is proportionate to a specific assurance objective.\footnote{Google DeepMind, ``\href{https://storage.googleapis.com/deepmind-media/gemini/gemini_4_safety_report.pdf}{Gemini 4 Argon Safety \& Security Report}.''} Anthropic has suggested that it would ``generally seek'' to share information about model architectures with external evaluators if it is ``highly relevant'' to risk assessments, while noting that ``information about architectures presents a high risk of inadvertent leaks.''\footnote{Anthropic, ``\href{https://www.anthropic.com/news/an-update-on-independent-evaluators}{An update on independent evaluators}.''}

Even where developers do not provide such access, they can still report the other indicators---though it would then be harder for third parties to verify the results since they could not run the underlying tests on the CoT themselves.

\Needspace{7\baselineskip}

\begin{indicatorpanel}

\textbf{Third-Party CoT Access and Base Architecture Specifications}

\textbf{Policy Recommendation:} Developers should provide external access to raw reasoning traces and relevant base architecture specifications.

\textbf{Rationale:} Third-party access enables initial warning signs about whether models continue to use CoT and if CoT is necessary for the model's complex reasoning capabilities. Access allows independent parties to verify the claims made in system cards with respect to the subsequent disclosures proposed here.

\textbf{Disclosure:} API-level access to raw, unsummarised reasoning traces for qualified third parties (such as embedded evaluators, Independent Verification Organizations, governments, accredited researchers) and details of relevant base architecture specifications that they can independently validate, with clear terms specifying the conditions of access.

\end{indicatorpanel}

\Needspace{6\baselineskip}

\subsection{The CoT Must Be in Readable Human Language}\label{the-cot-must-be-in-readable-human-language}

Even when CoT exists, it can drift from readable human language into compressed, model-convenient text that humans struggle to parse. Schoen et al. observe frontier reasoning models departing from standard English, using tokens in ways that are unnatural to human readers.\footnote{Schoen et al., ``\href{https://arxiv.org/abs/2509.15541}{Stress Testing Deliberative Alignment for Anti-Scheming Training}.''} For example, in a recent OpenAI report about a case of misaligned agent activity, ``translations'' marked as ``plain-language interpretations'' of relevant examples of sub-agents' CoT were provided alongside the original, presumably due to the latter's perceived illegibility to the reader.\footnote{OpenAI, ``\href{https://alignment.openai.com/misalignment-reports/suppressing-misalignment-monitoring-alerts-and-disabling-misalignment-monitoring}{Suppressing misalignment monitoring alerts and disabling misalignment monitoring}.''} This is a surface-visible failure: the non-standard text is immediately apparent on inspection, though its extent is difficult to quantify without a dedicated measure.

If the CoT uses this unnatural language, the ability to effectively monitor the model is reduced. For example, if a model is reasoning about scheming, but doing so in an uninterpretable way, it is more difficult to notice this scheming behavior.\footnote{Meinke et al., ``\href{https://arxiv.org/abs/2412.04984}{Frontier Models are Capable of In-context Scheming}.''} Therefore, we propose rating legibility on a fixed rubric, from unintelligible (0) to easy to follow (4), using a large language model (LLM) grader calibrated against human ratings.\footnote{Emmons et al., ``\href{https://arxiv.org/abs/2507.05246}{When Chain of Thought is Necessary, Language Models Struggle to Evade Monitors}.''} The rating concerns readability only, not whether the reasoning is correct or faithful, which are tackled in separate indicators.

\Needspace{7\baselineskip}

\begin{indicatorpanel}

\textbf{Naturalness}

\textbf{Diagnostic Question:} Is the CoT still in readable human language?

\textbf{Rationale:} To use the CoT to monitor harmful actions, monitors must be able to extract semantic meaning from it. Schoen et al. show that CoT may be drifting away from readable human language;\footnote{Schoen et al. ``\href{https://arxiv.org/abs/2509.15541}{Stress Testing Deliberative Alignment for Anti-Scheming Training}.''} this indicator measures whether and how far.

\textbf{Disclosure:} Mean legibility score on a 0--4 rubric assigned by an LLM grader calibrated against human ratings on CoTs from tasks the model solves, reported across task types.\footnote{Emmons et al., ``\href{https://arxiv.org/abs/2507.05246}{When Chain of Thought is Necessary, Language Models Struggle to Evade Monitors}.''} The rubric scores whether a human with the same language capabilities as the user can follow the reasoning without external aid.

\textbf{Current Practice}: Anthropic, OpenAI, and Google DeepMind report qualitative and quantitative evidence of CoT legibility, with Google DeepMind reporting legibility in line with our proposed disclosure metric.\footnote{Google DeepMind, ``\href{https://storage.googleapis.com/deepmind-media/gemini/gemini_4_safety_report.pdf}{Gemini 4 Argon Safety \& Security Report}.''} However, the developers differ in their operational definitions of legibility and the evaluations and metrics used to measure faithfulness.\footnote{See for example: Anthropic, ``\href{https://www-cdn.anthropic.com/0339e6a7c5c7b87f5c07798616dc32c215d14235/Claude\%20Fable\%205.1\%20&\%20Claude\%20Mythos\%205.1\%20System\%20Card.pdf}{System Card: Claude Fable 5.1 \& Claude Mythos 5.1}''; Anthropic. ``\href{https://www-cdn.anthropic.com/fc1b44717c85dc068bc6ba5024219938094694bd/Claude\%20Opus\%205.5\%20System\%20Card.pdf}{System Card: Claude Opus 5.5}''; OpenAI, ``\href{https://deploymentsafety.openai.com/gpt-6-astra}{GPT-6 Astra System Card}''; Google DeepMind, ``\href{https://storage.googleapis.com/deepmind-media/gemini/gemini_4_safety_report.pdf}{Gemini 4 Argon Safety \& Security Report}.''}

\end{indicatorpanel}

\Needspace{6\baselineskip}

\subsection{The CoT Must Be Faithful}\label{the-cot-must-be-faithful}

A readable CoT is only useful if it reflects the computation the model actually performed and, as such, is faithful.\footnote{Lanham et al., ``\href{https://arxiv.org/abs/2307.13702}{Measuring Faithfulness in Chain-of-Thought Reasoning}''; Chen et al., ``\href{https://arxiv.org/abs/2505.05410}{Reasoning Models Don\textquotesingle t Always Say What They Think}''; Barez et al., ``\href{https://aigi.ox.ac.uk/publications/chain-of-thought-is-not-explainability/}{Chain-of-Thought Is Not Explainability}.''} A model's CoT can be unfaithful in three ways.\footnote{Liu et al., ``\href{https://arxiv.org/abs/2602.13904}{Diagnosing Pathological Chain-of-Thought in Reasoning Models}.''} First, post-hoc rationalization: the model produces plausible-sounding justifications for an answer already determined after the prompt, so the reasoning that appears is reconstructed from the answer rather than producing it.\footnote{Turpin et al., ``\href{https://arxiv.org/abs/2305.04388}{Language Models Don\textquotesingle t Always Say What They Think: Unfaithful Explanations in Chain-of-Thought Prompting}.''} Second, the model may genuinely need the forward-pass computation of the CoT to arrive at its answer, but the emitted tokens do not reflect that computation.\footnote{Pfau et al., ``\href{https://arxiv.org/abs/2404.15758}{Let's Think Dot by Dot: Hidden Computation in Transformer Language Models}.''} Here, the text acts as scaffolding for internal computation rather than a window onto it. In both cases, a monitor reading the CoT risks being misled about the reasoning of the model. Third, a model may hide its true reasoning in the CoT through steganography.\footnote{Roger and Greenblatt, ``\href{https://arxiv.org/abs/2310.18512}{Preventing Language Models From Hiding Their Reasoning}; Skaf et al. \href{https://arxiv.org/abs/2506.01926}{Large language models can learn and generalize steganographic chain-of-thought under process supervision}.''} The CoT may carry load-bearing content in specific word choices, token orderings, or statistical patterns that are invisible to natural reading but carry computational content.

There is already evidence that suggests unfaithful CoT is a present reality, including among the most advanced publicly available models. For example, Anthropic's latest system cards include unfaithful thinking scores as part of an ``Automated Behavioral Audit'' suite, which shows consistent presence of unfaithful thinking, although this trend is improving in internal evaluations of more powerful models.\footnote{Anthropic, ``\href{https://www-cdn.anthropic.com/870c8f525702625d2c62fc6dd04c857e3250bec1/Claude\%20Sonnet\%205.5\%20System\%20Card.pdf}{System Card: Claude Sonnet 5.5}.''} OpenAI's GPT-6 Astra system card also suggested reduced CoT faithfulness in some scenarios compared with a previous-generation model.\footnote{OpenAI, ``\href{https://deploymentsafety.openai.com/gpt-6-astra}{GPT-6 Astra System Card}.''}

No single evaluation catches all three.\footnote{Barez et al., ``\href{https://aigi.ox.ac.uk/publications/chain-of-thought-is-not-explainability/}{Chain-of-Thought Is Not Explainability}.''} Cue-based tests probe post-hoc rationalization by checking whether the CoT acknowledges factors that demonstrably influenced the answer.\footnote{Turpin et al., ``\href{https://arxiv.org/abs/2305.04388}{Language Models Don\textquotesingle t Always Say What They Think: Unfaithful Explanations in Chain-of-Thought Prompting}.''} Perturbation and necessity tests probe scaffolding by intervening on the CoT and checking whether performance degrades as expected.\footnote{Meek et al., ``\href{https://arxiv.org/abs/2510.27378}{Measuring Chain-of-Thought Monitorability Through Faithfulness and Verbosity}.''} Paraphrasing tests probe for steganography by rewording the CoT while preserving apparent meaning and checking whether task performance drops;\footnote{Lanham et al., ``\href{https://arxiv.org/abs/2307.13702}{Measuring Faithfulness in Chain-of-Thought Reasoning}.''} a drop indicates the exact wording was doing computational work invisible to natural reading.

Because degradation can take any of the three forms above, and no single test detects all of them, we propose disclosure against a suite of faithfulness tests, each targeting a specific failure mode, rather than a single headline score.

Another consideration for CoT faithfulness is controllability, involving models ``deliberately reshaping or obscuring their reasoning when they know they are being monitored''.\footnote{OpenAI, ``\href{https://deploymentsafety.openai.com/gpt-6-astra}{GPT-6 Astra System Card}.''} Yueh-Han et al. suggested models struggle to do this;\footnote{Yueh-Han et al., ``\href{https://arxiv.org/abs/2603.05706}{Reasoning Models Struggle to Control their Chains of Thought}.''} however, in September 2026, OpenAI reported that GPT-6 Astra's ability to control its CoT is ``substantially higher than that of GPT-5.6 Sol and GPT-5.5 Thinking,''\footnote{OpenAI, ``\href{https://deploymentsafety.openai.com/gpt-6-astra}{GPT-6 Astra System Card}.''} and Anthropic reported that Claude Mythos 5.1 showed ``slightly increased chain-of-thought controllability with the best-performing prompt'' compared to Claude Mythos Preview.\footnote{Anthropic, ``\href{https://www-cdn.anthropic.com/0339e6a7c5c7b87f5c07798616dc32c215d14235/Claude\%20Fable\%205.1\%20&\%20Claude\%20Mythos\%205.1\%20System\%20Card.pdf}{System Card: Claude Fable 5.1 \& Claude Mythos 5.1}.''} Google DeepMind assessed that Gemini 4 Argon's CoT controllability is no higher than that of a previous family model---Gemini 3.1 Pro---when controlling for CoT length.\footnote{Google DeepMind, ``\href{https://storage.googleapis.com/deepmind-media/gemini/gemini_4_safety_report.pdf}{Gemini 4 Argon Safety \& Security Report}.''}

Importantly, a model's ability to control its CoT does not guarantee it is unfaithful, especially because the model still needs to deliberately leverage the capability to obfuscate what is in its CoT while not undermining its capability to complete the task, and it may decide not to do so. However, CoT controllability is likely a strong prerequisite for models to actively undermine the faithfulness of their CoT.\footnote{OpenAI, ``\href{https://deploymentsafety.openai.com/gpt-6-astra}{GPT-6 Astra System Card}.''} We recommend ongoing tracking and disclosure of CoT controllability indicators on an expanded basis by developers as a complement to this framework, including how specific prompting techniques influence eliciting these capabilities and whether models with greater ability to control their CoT at a given context length are more likely to do so in adversarial settings.

\Needspace{7\baselineskip}

\begin{indicatorpanel}

\textbf{Faithfulness}

\textbf{Diagnostic Question:} Does the CoT reflect the model's actual reasoning?

\textbf{Rationale:} Indicates whether the text in the CoT corresponds to the computation the model actually performed. Low faithfulness can arise from post-hoc rationalization, from the CoT acting as scaffolding rather than a window on computation, or from steganographic encoding. A monitor reading the CoT is misled in all three cases. Comparing scores before and after post-training reveals whether post-training has degraded faithfulness.\footnote{Baker et al., ``\href{https://arxiv.org/abs/2503.11926}{Monitoring Reasoning Models for Misbehavior and the Risks of Promoting Obfuscation}''; Guan et al., ``\href{https://arxiv.org/abs/2512.18311}{Monitoring Monitorability}.''}

\textbf{Disclosure:} Scores on a faithfulness suite that covers all three failure modes, such as cue-based tests for post-hoc rationalization, perturbation and necessity tests for scaffolding, and paraphrasing tests for steganographic encoding. Reported both pre- and post-training at each major model release, across multiple domains, of the kinds described in Liu et al., Meek et al., and Mayne et al.\footnote{Liu et al., ``\href{https://arxiv.org/abs/2602.13904}{Diagnosing Pathological Chain-of-Thought in Reasoning Models}''; Meek et al., ``\href{https://arxiv.org/abs/2510.27378}{Measuring Chain-of-Thought Monitorability Through Faithfulness and Verbosity}''; Mayne et al., ``\href{https://arxiv.org/abs/2602.02639}{A Positive Case for Faithfulness: LLM Self-Explanations Help Predict Model Behavior}.''}

\textbf{Current Practice:} No leading developer reports a comprehensive faithfulness suite. Anthropic and OpenAI report evaluations of CoT faithfulness but differ in how they evaluate and report it.\footnote{See for example Anthropic, ``\href{https://www-cdn.anthropic.com/870c8f525702625d2c62fc6dd04c857e3250bec1/Claude\%20Sonnet\%205.5\%20System\%20Card.pdf}{System Card: Claude Sonnet 5.5}''; OpenAI, ``\href{https://deploymentsafety.openai.com/gpt-6-astra}{GPT-6 Astra System Card}.''} Google DeepMind does not systematically report faithfulness evaluations in its frontier safety framework reports; however, it has engaged with this topic in separate research reports.\footnote{Emmons et al., ``\href{https://arxiv.org/abs/2507.05246}{When Chain of Thought is Necessary, Language Models Struggle to Evade Monitors}.''} Mayne et al. externally evaluated 18 frontier models, demonstrating a potential cross-model evaluation approach.\footnote{Mayne et al., ``\href{https://arxiv.org/abs/2602.02639}{A Positive Case for Faithfulness: LLM Self-Explanations Help Predict Model Behavior}.''}

\end{indicatorpanel}

\Needspace{6\baselineskip}

\subsection{The Monitor Must Catch Harmful Behavior}\label{the-monitor-must-catch-harmful-behavior}

A readable and faithful CoT is only useful for monitorability if a monitor can actually use it to detect dangerous behavior. This depends on the monitor's capability,\footnote{Baker et al., ``\href{https://arxiv.org/abs/2503.11926}{Monitoring Reasoning Models for Misbehavior and the Risks of Promoting Obfuscation}''; Guan et al., ``\href{https://arxiv.org/abs/2512.18311}{Monitoring Monitorability}.''} its freedom from collusion with the monitored model,\footnote{Greenblatt et al., ``\href{https://arxiv.org/abs/2312.06942}{AI Control: Improving Safety Despite Intentional Subversion}''; Gardner-Challis et al., ``\href{https://arxiv.org/abs/2602.20628}{When can we trust untrusted monitoring? A safety case sketch across collusion strategies}.''} and its robustness to rationalization or biased reasoning in the CoT, which could result in the monitor being misled even without a deliberate attempt to do so.\footnote{Anthropic, ``\href{https://www.anthropic.com/research/alignment-assessment-cybersecurity-incidents\#which-monitors-would-have-caught-this-behavior}{An alignment assessment of recent cybersecurity incidents}''; METR. ``\href{https://metr.org/blog/2026-10-20-misalignment-risk-report-august-2026-review/}{Review of the `Misalignment in high-stakes settings' section in Anthropic's risk report (August 2026)}.''} Monitor detection rate is the bottom-line lagging indicator: a model can score well on every leading CoT-quality indicator while still evading monitors.\footnote{Arnav et al., ``\href{https://arxiv.org/abs/2505.23575}{CoT Red-Handed: Stress Testing Chain-of-Thought Monitoring}.''} Questions about whether the monitor is operating over all relevant traffic are out of scope of this report, but we note that weaknesses in monitor coverage have been identified in red-teaming exercises.\footnote{See for example: METR, ``\href{https://metr.org/blog/2026-10-22-red-teaming-openai-agent-monitoring/}{Red-Teaming OpenAI's Internal Agent Monitoring Systems}.''}

\Needspace{7\baselineskip}

\begin{indicatorpanel}

\textbf{Detection}

\textbf{Diagnostic Question:} Does the monitoring system catch dangerous behavior?

\textbf{Rationale:} Reveals what proportion of harmful behaviors evade CoT-based monitoring in standardized evaluations. A model can score well on Naturalness and Faithfulness while still evading monitors through strategies those tests do not capture.

\textbf{Disclosure:} Proportion of harmful behaviors evading the monitoring system across standardized evaluation scenarios, with a shared set of threat archetypes for cross-developer comparison. Disclosures should identify material differences between the evaluated monitoring setup and the production version, with particular reference to any limitations of the former that could impact the level of confidence around claims about the robustness of the latter.

\textbf{Current Practice:} Building on Guan et al., OpenAI reports aggregate performance rates for monitors in non-adversarial conditions across different monitor access setups and separately reports individual scores in various adversarial conditions.\footnote{See for example: OpenAI, ``\href{https://deploymentsafety.openai.com/gpt-6-astra}{GPT-6 Astra System Card}''; OpenAI, ``\href{https://deploymentsafety.openai.com/gpt-5-6}{GPT-5.6 System Card}.''} Anthropic reports monitor performance across a range of adversarial datasets, alongside SHADE-Arena and LinuxArena stealth evaluation results.\footnote{See for example: Anthropic, ``\href{https://www-cdn.anthropic.com/0339e6a7c5c7b87f5c07798616dc32c215d14235/Claude\%20Fable\%205.1\%20&\%20Claude\%20Mythos\%205.1\%20System\%20Card.pdf}{System Card: Claude Fable 5.1 \& Claude Mythos 5.1}''; Anthropic, ``\href{https://www-cdn.anthropic.com/870c8f525702625d2c62fc6dd04c857e3250bec1/Claude\%20Sonnet\%205.5\%20System\%20Card.pdf}{System Card: Claude Sonnet 5.5}''; Kusatov et al., ``\href{https://arxiv.org/abs/2506.15740}{SHADE-Arena: Evaluating Sabotage and Monitoring in LLM Agents}''; Tracy et al., ``\href{https://arxiv.org/abs/2604.15384}{LinuxArena: A Control Setting for AI Agents in Live Production Software Environments}.''} Google DeepMind reports detection metrics in setups including LinuxArena, an internal codebase version, and honeypots.\footnote{Google DeepMind, ``\href{https://storage.googleapis.com/deepmind-media/gemini/gemini_4_safety_report.pdf}{Gemini 4 Argon Safety \& Security Report}''; Guan et al., ``\href{https://arxiv.org/abs/2512.18311}{Monitoring Monitorability}''; Tracy et al., ``\href{https://arxiv.org/abs/2604.15384}{LinuxArena: A Control Setting for AI Agents in Live Production Software Environments}.''} METR's SHUSHCAST framework provides an independent benchmark.\footnote{Kinniment et al. ``\href{https://metr.org/blog/2026-01-19-early-work-on-monitorability-evaluations/\#introduction}{Early work on monitorability evaluations}''; METR, ``\href{https://metr.org/blog/2026-05-19-frontier-risk-report/\#executive-summary-and-guide-to-the-report}{Frontier Risk Report (February to March 2026)}.''} No standardized benchmark and configuration exist for cross-model and -developer comparisons, but these demonstrate that measurement is practical.

\end{indicatorpanel}

\Needspace{6\baselineskip}

\subsection{CoT Must Add Value Over Action-Only Monitoring}\label{cot-must-add-value-over-action-only-monitoring}

Finally, even a capable monitor only justifies effort to preserve CoT when CoT access materially improves its performance over observing actions alone.\footnote{Arnav et al., ``\href{https://arxiv.org/abs/2505.23575}{CoT Red-Handed: Stress Testing Chain-of-Thought Monitoring}''; Guan et al., ``\href{https://arxiv.org/abs/2512.18311}{Monitoring Monitorability}.''} Given failure of any earlier framework condition makes the later conditions moot, if the value gap shrinks at this final stage, either the quality of CoT has become too poor to help or action-only methods have improved, meaning that efforts focusing on CoT are less important. In either case, this information would be useful to disclose, and the priority placed on preserving CoT should be revisited in such circumstances.

\Needspace{7\baselineskip}

\begin{indicatorpanel}

\textbf{Uplift}

\textbf{Diagnostic Question}: How much does CoT access improve monitoring?

\textbf{Rationale:} Indicates the performance gap between monitors that can read the CoT and monitors that can only observe actions. This quantifies how much value CoT access actually provides for safety monitoring.

\textbf{Disclosure:} Monitoring effectiveness under three conditions (action-only, CoT-only, and full-context)\footnote{OpenAI, ``\href{https://deploymentsafety.openai.com/gpt-6-astra}{GPT-6 Astra System Card}.''} across standardized threat scenarios, including one that involves biased reasoning in the CoT. Improvements and reductions in detection---disaggregated by threat scenario---should be reported together with an aggregated uplift score.

\textbf{Current Practice:} OpenAI's GPT-5.4 Thinking system card\footnote{OpenAI, ``\href{https://deploymentsafety.openai.com/gpt-5-4-thinking}{GPT-5.4 Thinking System Card}.''} was the first to report this as part of a standard model release; subsequent system cards have continued this, including for GPT-5.6 and GPT-6 Astra.\footnote{OpenAI. ``\href{https://deploymentsafety.openai.com/gpt-5-6}{GPT-5.6 System Card}''; OpenAI, ``\href{https://deploymentsafety.openai.com/gpt-6-astra}{GPT-6 Astra System Card}.''} Anthropic publishes comparisons with and without reasoning access on its SHADE-Arena stealth evaluation.\footnote{See for example: Anthropic, ``\href{https://www-cdn.anthropic.com/0339e6a7c5c7b87f5c07798616dc32c215d14235/Claude\%20Fable\%205.1\%20&\%20Claude\%20Mythos\%205.1\%20System\%20Card.pdf}{System Card: Claude Fable 5.1 \& Claude Mythos 5.1}''; Anthropic. ``\href{https://www-cdn.anthropic.com/fc1b44717c85dc068bc6ba5024219938094694bd/Claude\%20Opus\%205.5\%20System\%20Card.pdf}{System Card: Claude Opus 5.5}''; Kutasov et al., ``\href{https://arxiv.org/abs/2506.15740}{SHADE-Arena: Evaluating Sabotage and Monitoring in LLM Agents}.''} Google DeepMind reports monitoring effectiveness results across action-only, CoT-only, and combined monitors.\footnote{Google DeepMind, ``\href{https://storage.googleapis.com/deepmind-media/gemini/gemini_4_safety_report.pdf}{Gemini 4 Argon Safety \& Security Report}.''} In general, these disclosures differ in task coverage, experimental conditions, and metrics, limiting direct comparisons of uplift across developers.

\end{indicatorpanel}

\Needspace{6\baselineskip}

\section{3 \textbar{} Method for Identifying Monitorability Indicators}\label{method-for-identifying-monitorability-indicators}

\Needspace{6\baselineskip}

\subsection{Generating Candidate Indicators}\label{generating-candidate-indicators}

We compiled 21 candidate transparency disclosures that would be informative about whether chain-of-thought (CoT) monitorability is being preserved. These were generated through two complementary processes. First, we conducted a review of recent CoT monitorability research literature, developer disclosures, and other sources (see {\begingroup\hypersetup{linkcolor=MainTextLinkColor,urlcolor=MainTextLinkColor}\hyperref[sources-reviewed]{Appendix, Sources Reviewed}\endgroup{}}). Then, we mapped the framework of CoT monitorability in {\begingroup\hypersetup{linkcolor=MainTextLinkColor,urlcolor=MainTextLinkColor}\hyperref[a-framework-for-chain-of-thought-monitoring]{Section 2}\endgroup{}} to specific indicators that would reveal information about the conditions that need to hold to preserve CoT monitorability.

\Needspace{6\baselineskip}

\subsection{Scoring}\label{scoring}

Each candidate disclosure was scored from 1--5 (1=low, 5=high) on five dimensions.

\begin{itemize}
\item
  \textbf{Informativeness}. How much information would this indicator give about the CoT monitorability of the system in question? A measure scoring 5 is directly predictive of the level of degradation; a measure scoring 1 is minimally informative.
\item
  \textbf{Technology Readiness}. Does a validated measurement method already exist? A score of 1 indicates no methodology exists and a score of 5 means there is a validated, standardized method available.
\item
  \textbf{External Verifiability.} Assuming that the technology exists, can a third party verify the indicator result without trusting the developer? Developers may have incentives not to report accurate numbers and, therefore, external verifiability increases confidence in disclosed results. A score of 1 means verification is completely trust-dependent and a score of 5 means the result is fully externally verifiable.
\item
  \textbf{Developer Tractability.} Assuming that the technology exists, how easily can the developer collect and report this indicator from a technical perspective? What are the infrastructure requirements and technical effort required to obtain this information? A score of 1 indicates major changes and high technical effort would be required and a score of 5 means the information is already collected and easy to report.
\item
  \textbf{Implementation Feasibility}. How realistic is it that developers would adopt this disclosure from a strategic perspective? We examine the intellectual property (IP) sensitivity of the information and the alignment with existing reporting practices. A score of 1 means there is substantial IP risk and more information required to be provided and a score of 5 means there is no IP risk and the information is already being provided by some developers).
\end{itemize}

The scores were assigned by the first author and reviewed by the second author in the original paper.\footnote{Davies and Guest, ``\href{https://openreview.net/forum?id=H5FSdlrdZs}{Transparency Disclosures for Chain-of-Thought Monitorability}.''} Prior to the original paper being published, approximately ten researchers with expertise in AI safety (including researchers at frontier labs), AI governance, and AI evaluations were consulted. The top-scoring indicators were presented to the researchers, and they were asked whether the scores were reasonable and whether we had missed any important potential indicators or considerations. The most substantive feedback was to avoid prescribing specific evaluation methods and instead frame disclosures in terms of the information to be reported, leaving developers flexibility in methodology. This shaped the final indicator descriptions in {\begingroup\hypersetup{linkcolor=MainTextLinkColor,urlcolor=MainTextLinkColor}\hyperref[a-framework-for-chain-of-thought-monitoring]{Section 2}\endgroup{}}.

\Needspace{6\baselineskip}

\subsection{Prioritization}\label{prioritization}

We selected at least one indicator for each of the five conditions in our framework ({\begingroup\hypersetup{linkcolor=MainTextLinkColor,urlcolor=MainTextLinkColor}\hyperref[a-framework-for-chain-of-thought-monitoring]{Section 2}\endgroup{}}). Grouping the candidates by the condition they bear on (see {\begingroup\hypersetup{linkcolor=MainTextLinkColor,urlcolor=MainTextLinkColor}\hyperref[full-scored-framework]{Appendix, Full Scored Framework}\endgroup{}}) let us choose the best indicator(s) within each condition rather than imposing a single threshold score across conditions. This means some indicators that were not selected received higher scores than those that were selected in other categories, because we were interested in the top-scoring indicator in each condition rather than comparing scores across conditions.

Within each condition, we chose the highest-scoring candidate as that condition's indicator, with three exceptions noted below.

First, for two conditions, the leading candidates tied on overall score. We broke the ties in favor of the more direct measure---the one that more strongly observes the condition itself rather than proxies. For readability (condition 2), naturalness and CoT coverage/verbosity both scored 18. We selected naturalness, which scores readability directly, over verbosity, which is a proxy. For marginal value (condition 5), uplift and inference compute per reasoning token both scored 16. We selected uplift, which directly measures how much CoT access improves monitoring, over inference compute, which is a proxy.

Second, the CoT existence and necessity condition (condition 1) contains two indicators to understand whether the CoT is load-bearing for the model's complex reasoning capabilities. As discussed in {\begingroup\hypersetup{linkcolor=MainTextLinkColor,urlcolor=MainTextLinkColor}\hyperref[the-cot-must-exist-and-be-necessary-for-complex-reasoning]{Section 2 (The CoT Must Exist and Be Necessary For Complex Reasoning)}\endgroup{}}, models may incorporate latent reasoning architectures like looped transformers, which could result in lower reliance on the CoT for reasoning even if it still exists. To mitigate the risk of the presence of the CoT creating a false sense of security about model monitorability, we recommend relevant base architecture specifications be disclosed---alongside CoT access---for qualified third parties to independently validate. While the former indicator scores lower than the latter, we believe disclosing both is essential to robustly fulfill this condition.

Third, the faithfulness indicator (condition 3) is a suite rather than a single measure. Faithfulness fails in three distinct ways---post-hoc rationalization, scaffolding, and steganography---and, as {\begingroup\hypersetup{linkcolor=MainTextLinkColor,urlcolor=MainTextLinkColor}\hyperref[a-framework-for-chain-of-thought-monitoring]{Section 2}\endgroup{}} argues, no single test detects all three. The indicator is therefore the set of complementary tests that together cover the three modes: cue-based tests for rationalization, perturbation and necessity tests for scaffolding, and paraphrasing tests for steganographic encoding. We select these for their coverage of the failure modes rather than by score, though each also scores well among the faithfulness candidates.

The full scoring for all 21 candidates, with justifications for each score, is in {\begingroup\hypersetup{linkcolor=MainTextLinkColor,urlcolor=MainTextLinkColor}\hyperref[full-scored-framework]{Appendix (Full Scored Framework)}\endgroup{}}.

\Needspace{6\baselineskip}

\subsection{Limitations}\label{limitations}

Two limitations are worth flagging. First, the rubric in {\begingroup\hypersetup{linkcolor=MainTextLinkColor,urlcolor=MainTextLinkColor}\hyperref[scoring]{Section 3 (Scoring)}\endgroup{}} anchors only the endpoints (1 and 5); the intermediate values 2, 3, and 4 are not formally defined, so scores in the middle of the range reflect author judgment rather than a shared standard. Selecting the best indicator within each condition reduces our reliance on these midpoints, but they still affect borderline within-condition comparisons. Second, scores were assigned by two authors and validated through consultation with external researchers rather than by independent scoring from multiple raters with inter-rater reliability analysis. The precise rankings, and the borderline cases, should therefore be treated as indicative rather than definitive.

\Needspace{6\baselineskip}

\section{4 \textbar{} Cross-Developer Audit of Current Disclosures}\label{cross-developer-audit-of-current-disclosures}

We audited Anthropic, Google DeepMind, and OpenAI---arguably the leading three frontier developers---against our five indicators. This audit was based on the most recent system cards and published research from each of the three developers (Figure 3).\footnote{We looked at all published research from the last year with any reference to CoT.}

\begin{figure}[htbp]\centering

\includegraphics[width=\linewidth]{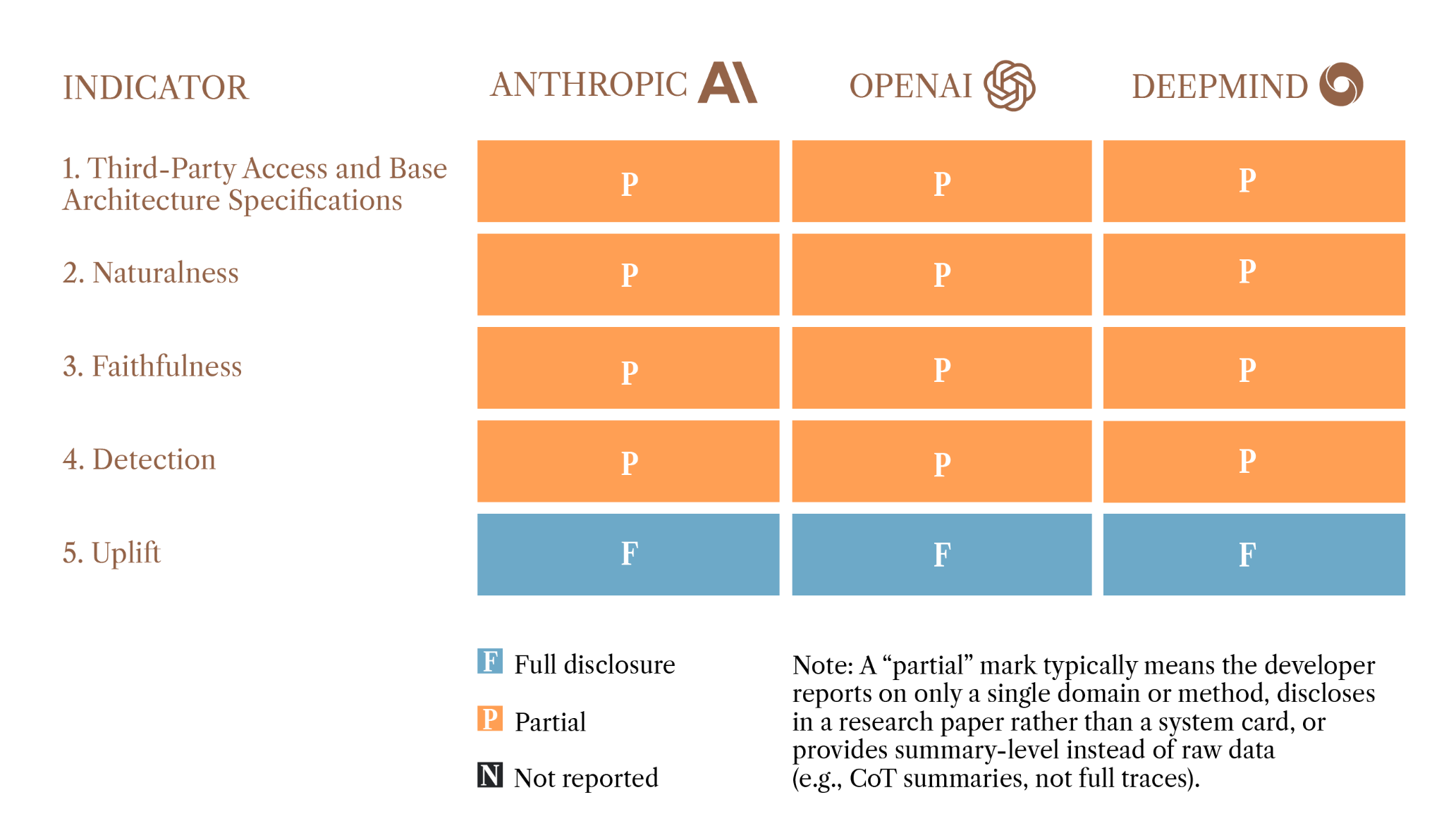}

\caption{Cross-Developer Audit of Disclosure Against the Five Priority Indicators}\label{fig:3}

\end{figure}

\textbf{No developer covers the full set, but every indicator has been reported by at least one company}. This suggests that the full set is feasible to adopt, though challenges around standardization, coverage, and verification remain. We recommend repeating this audit with deeper access than public disclosures allow, either by embedded evaluators working within developers or by an Independent Verification Organization.

The most notable disclosure was OpenAI's reporting of chain-of-thought (CoT) monitoring uplift in the GPT-5.4 Thinking System Card,\footnote{OpenAI, ``\href{https://deploymentsafety.openai.com/gpt-5-4-thinking}{GPT-5.4 Thinking System Card}.''} which was the first time any developer has reported this as part of a standard model release. Further such disclosures were made, together with reports of a notable decline in CoT monitorability, for GPT-6 Astra.\footnote{OpenAI, ``\href{https://deploymentsafety.openai.com/gpt-5-6}{GPT-5.6 System Card}''; OpenAI, ``\href{https://deploymentsafety.openai.com/gpt-6-astra}{GPT-6 Astra System Card}.''} Anthropic also reports scores on uplift, faithfulness, and legibility alongside monitor performance, though it defines and operationalizes these concepts differently, underscoring how a lack of standardization can frustrate cross-model and cross-developer comparisons.\footnote{See for example: Anthropic, ``\href{https://www-cdn.anthropic.com/0339e6a7c5c7b87f5c07798616dc32c215d14235/Claude\%20Fable\%205.1\%20&\%20Claude\%20Mythos\%205.1\%20System\%20Card.pdf}{System Card: Claude Fable 5.1 \& Claude Mythos 5.1}''; Anthropic, ``\href{https://www-cdn.anthropic.com/870c8f525702625d2c62fc6dd04c857e3250bec1/Claude\%20Sonnet\%205.5\%20System\%20Card.pdf}{System Card: Claude Sonnet 5.5}.''} Google DeepMind has disclosed detailed CoT monitorability evaluation results for Gemini 4 Argon, including on legibility, monitor performance, and uplift.\footnote{Google DeepMind, ``\href{https://storage.googleapis.com/deepmind-media/gemini/gemini_4_safety_report.pdf}{Gemini 4 Argon Safety \& Security Report}.''}

\Needspace{6\baselineskip}

\section{5 \textbar{} Discussion}\label{discussion}

\textbf{Direct Detection vs. Causal Mechanism Measures.} There are indicators that measure poor monitorability (direct detection measures such as faithfulness scores, bypass rate, and encoding detection) and indicators that track factors that cause poor monitorability (causal mechanism measures such as optimization pressure and reinforcement learning from human feedback effects on chain of thought (CoT). The five priority indicators are predominantly direct detection measures, which we believe are the most actionable for transparency disclosures. They are more externally verifiable, easier for developers to report, and more directly informative about the current state of monitorability. Causal mechanism measures, while valuable for understanding why monitorability degrades, are often unverifiable externally (e.g., optimization pressure on CoT). Furthermore, frontier developers are much less likely to disclose causal indicators, as it involves disclosing information about their training setup, which is highly commercially sensitive.

\textbf{Other Kinds of Indicators.} The indicators discussed in this report are technical and could apply to an individual AI system. But it might also be valuable for frontier developers to disclose other kinds of information that give insight into how well the developer is doing on preserving CoT monitorability, or how much effort the developer might be willing to spend in the future on preserving monitorability. For example, frontier developers could potentially disclose institutional details, such as what internal policies a developer has about monitorability. Information developers already disclose publicly, such as any commitments to monitorability in Frontier AI Safety Policies or the compliance and governance frameworks they have in place to meet legal requirements, could also be useful as an indicator.

\textbf{Disclosure Does Not Guarantee Preservation.} Transparency disclosures make degradation visible, but they do not by themselves prevent it. A developer could report declining faithfulness scores and take no corrective action. That said, visibility would have two advantages, both mentioned in {\begingroup\hypersetup{linkcolor=MainTextLinkColor,urlcolor=MainTextLinkColor}\hyperref[the-value-of-transparency-about-monitorability-indicators]{Section 1}\endgroup{}}. First, making the problem more visible would allow others to respond to it, such as with technical or policy efforts to more strongly facilitate CoT monitorability. Second, visibility would strengthen frontier developers' incentives to preserve monitorability. For example, visibility would make it clearer whether frontier developers are defecting from norms around preserving monitorability.

\textbf{Gaming and Goodhart Risks.} Any disclosure framework faces the risk that developers optimize for the reported metric rather than the underlying property. We have deliberately avoided prescribing specific benchmarks or evaluation methods partly for this reason: the indicators specify what information to disclose, not which test to run. This reduces the attack surface for gaming, though it does not eliminate it. The inclusion of multiple complementary indicators also provides some protection, as gaming faithfulness scores while neglecting monitor bypass rate, for example, would produce a visibly inconsistent disclosure profile.

\textbf{Reporting Cadence.} New models are coming out rapidly, meaning that, for the indicators to be informative, it will be important for frontier developers to regularly disclose them, such as with a disclosure each quarter or upon each model release.

\textbf{Relationship to Other Proposals.} Since this report was first circulated, other proposals on CoT monitorability disclosures have emerged. Greenblatt et al. have advocated for developers reporting externally verified information about ``opaque serial depth''---defined as ``the length of the longest computation that can be done without the use of interpretable intermediate steps like chain of thought''\footnote{Greenblatt et al., ``\href{https://blog.redwoodresearch.org/p/proposal-for-tracking-the-effects}{Proposal for tracking the effects of architecture on monitorability}.''}---across models; sharing third-party-reviewed information about the monitorability of their models across different architectures and training approaches together with the evidence used to support this; and publishing a policy that includes which architectures could degrade monitorability, the evidence that would be required to rule this out, and their approach to the trade-off between performance improvements and monitorability when researching different architectural approaches.\footnote{Brown-Cohen et al., ``\href{https://arxiv.org/abs/2603.09786}{Quantifying the Necessity of Chain of Thought through Opaque Serial Depth}.''} Shah and Dragan advocate for greater efforts to measure CoT transparency: preserving transparent architectures through developer commitments or potential regulation; subjecting model training practices to greater scrutiny to ensure they do not incentivize CoT opacity; and requiring developers to demonstrate that any deviation from CoT is not accompanied by degraded monitorability.\footnote{Shah and Dragan, ``\href{https://institute.deepmind.com/essays/the-case-for-reasoning-transparency/}{The case for reasoning transparency}.''} Our framework complements and builds on these proposals by providing monitorability indicators that integrate readily into existing system-card practices and are standardizable for more robust cross-model and cross-developer comparisons.

\Needspace{6\baselineskip}

\section{6 \textbar{} Conclusion}\label{conclusion}

There is significant consensus among frontier developers and researchers that chain-of-thought (CoT) monitorability is valuable and under threat. Transparency about monitorability indicators can help policymakers and researchers track this tool's effectiveness in overseeing frontier AI systems, inform technical and policy responses when evidence shows it is weakening, and create reputational and coordination-related incentives to preserve it. Transparency about five indicators seems particularly valuable and feasible: third-party CoT access and relevant base architecture specifications, naturalness, faithfulness, detection, and uplift.

Each of these indicators has been disclosed by at least one leading developer, showing that doing so is feasible. Frontier AI developers would ideally report these indicators as part of their existing system-card and safety-evaluation practices. Because current disclosures rely largely on developers' own evaluations, we also recommend that a cross-developer audit be repeated with deeper access than public disclosures allow, either by embedded evaluators working within developers or by an Independent Verification Organization.

\clearpage

\Needspace{6\baselineskip}

\section{Bibliography}\label{bibliography}

\begingroup\small\setlength{\parskip}{6pt}

\phantomsection\label{source-ref-87}

\hangindent=1.5em\hangafter=1\noindent Anthropic. ``Detecting and preventing distillation attacks.'' Anthropic News, February 23, 2026a. \url{https://www.anthropic.com/news/detecting-and-preventing-distillation-attacks}. Accessed: 2026-04-24.\par

\phantomsection\label{source-ref-88}

\hangindent=1.5em\hangafter=1\noindent Anthropic. ``System Card: Claude Opus 4.6.'' February 2026b. \url{https://www-cdn.anthropic.com/14e4fb01875d2a69f646fa5e574dea2b1c0ff7b5.pdf}.\par

\phantomsection\label{source-ref-89}

\hangindent=1.5em\hangafter=1\noindent Anthropic. ``System Card: Claude Fable 5 \& Claude Mythos 5'' June 9, 2026c. \url{https://www-cdn.anthropic.com/d00db56fa754a1b115b6dd7cb2e3c342ee809620.pdf}.\par

\phantomsection\label{source-ref-90}

\hangindent=1.5em\hangafter=1\noindent Anthropic. ``System Card: Claude Fable 5.1 \& Claude Mythos 5.1.'' September 1, 2026d. \url{https://www-cdn.anthropic.com/0339e6a7c5c7b87f5c07798616dc32c215d14235/Claude\%20Fable\%205.1\%20&\%20Claude\%20Mythos\%205.1\%20System\%20Card.pdf}.\par

\phantomsection\label{source-ref-91}

\hangindent=1.5em\hangafter=1\noindent Anthropic. ``An alignment assessment of recent cybersecurity incidents.'' September 9, 2026e.\url{https://www.anthropic.com/research/alignment-assessment-cybersecurity-incidents\#which-monitors-would-have-caught-this-behavior}. Accessed: 2026-10-22.\par

\phantomsection\label{source-ref-92}

\hangindent=1.5em\hangafter=1\noindent Anthropic. ``System Card: Claude Opus 5.5.'' September 22, 2026f. \url{https://www-cdn.anthropic.com/fc1b44717c85dc068bc6ba5024219938094694bd/Claude\%20Opus\%205.5\%20System\%20Card.pdf}.\par

\phantomsection\label{source-ref-93}

\hangindent=1.5em\hangafter=1\noindent Anthropic. ``System Card: Claude Sonnet 5.5.'' September 28, 2026g.\url{https://www-cdn.anthropic.com/870c8f525702625d2c62fc6dd04c857e3250bec1/Claude\%20Sonnet\%205.5\%20System\%20Card.pdf}.\par

\phantomsection\label{source-ref-94}

\hangindent=1.5em\hangafter=1\noindent Anthropic. ``An update on independent evaluators.'' October 15, 2026h.\url{https://www.anthropic.com/news/an-update-on-independent-evaluators}. Accessed: 2026-11-02.\par

\phantomsection\label{source-ref-95}

\hangindent=1.5em\hangafter=1\noindent Anwar, Usman, Bakker, T., Kianfar, D., Pinneri, C., Louizos, C. ``Analyzing and Improving Chain-of-Thought Monitorability Through Information Theory.''. arXiv, October 21, 2025. \url{https://arxiv.org/abs/2602.18297}.\par

\phantomsection\label{source-ref-96}

\hangindent=1.5em\hangafter=1\noindent Arnav, Benjamin, Bernabeu-Pérez, P., Helm-Burger, N., Kostolansky, T., Whittingham, H., and Phuong, M. ``CoT Red-Handed: Stress Testing Chain-of-Thought Monitoring.'' arXiv, November 24, 2025. \url{https://arxiv.org/abs/2505.23575}.\par

\phantomsection\label{source-ref-97}

\hangindent=1.5em\hangafter=1\noindent Baker, Bowen, Huizinga, J., Gao, L., Dou, Z., Guan, M.Y., Madry, A., Zaremba, W., Pachocki, J., and Farhi, D. ``Monitoring Reasoning Models for Misbehavior and the Risks of Promoting Obfuscation.''. arXiv, March 14, 2025. \url{https://arxiv.org/abs/2503.11926}.\par

\phantomsection\label{source-ref-98}

\hangindent=1.5em\hangafter=1\noindent Barez, Fazl., Wu, T.-Y., Arcuschin, I., Lan, M., Wang, V., Siegel, N., Collignon, N., Neo, C., Lee, I., Paren, A., Bibi, A., Trager, R., Fornasiere, D., Yan, J., Elazar, Y., and Bengio, Y. ``Chain-of-Thought Is Not Explainability.'' July 15, 2025. \url{https://aigi.ox.ac.uk/publications/chain-of-thought-is-not-explainability/}.\par

\phantomsection\label{source-ref-99}

\hangindent=1.5em\hangafter=1\noindent Brown-Cohen, Jonah, Lindner, D., and Shah, R. ``Quantifying the Necessity of Chain of Thought through Opaque Serial Depth.'', arXiv, March 10, 2026. \url{https://arxiv.org/abs/2603.09786}.\par

\phantomsection\label{source-ref-100}

\hangindent=1.5em\hangafter=1\noindent Chen, Yanda, Benton, J., Radhakrishnan, A., Uesato, J., Denison, C., Schulman, J., Somani, A., Hase, P., Wagner, M., Roger, F., Mikulik, V., Bowman, S. R., Leike, J., Kaplan, J., and Perez, E. ``Reasoning Models Don\textquotesingle t Always Say What They Think.'' May 8, 2025. \url{https://arxiv.org/abs/2505.05410}.\par

\phantomsection\label{source-ref-101}

\hangindent=1.5em\hangafter=1\noindent Davies, Jacob and Guest, O. ``Transparency Disclosures for Chain-of-Thought Monitorability.'' \emph{In} ICML 2026 Workshop on Technical AI Governance, July 12, 2026. \url{https://openreview.net/forum?id=H5FSdlrdZs}.\par

\phantomsection\label{source-ref-102}

\hangindent=1.5em\hangafter=1\noindent Delaney, Oscar, Guest, O., and Araujo, R. ``Policy Options for Preserving Chain of Thought Monitorability.'' Institute for AI Policy and Strategy (IAPS), September 24, 2025. \url{https://www.iaps.ai/research/policy-options-for-preserving-cot-monitorability}.\par

\phantomsection\label{source-ref-103}

\hangindent=1.5em\hangafter=1\noindent Emmons, Scott, Jenner, E., Elson, D. K., Saurous, R. A., Rajamanoharan, S., Chen, H., Shafkat, I., and Shah, R. ``When Chain of Thought is Necessary, Language Models Struggle to Evade Monitors.'' arXiv, July 7, 2025. \url{https://arxiv.org/abs/2507.05246}.\par

\phantomsection\label{source-ref-104}

\hangindent=1.5em\hangafter=1\noindent Ford, Celia. ``OpenAI's GPT-6 Astra might be too powerful to understand or control.'' Transformer, September 4, 2026. \url{https://www.transformernews.ai/p/openai-gpt-6-astra-might-be-too-powerful-to-understand-or-control}. Accessed: 2026-10-08.\par

\phantomsection\label{source-ref-105}

\hangindent=1.5em\hangafter=1\noindent Frontier Model Forum. ``Chain of Thought Monitorability.'' Issue Brief, January 27, 2026. \url{https://www.frontiermodelforum.org/issue-briefs/chain-of-thought-monitorability/}.\par

\phantomsection\label{source-ref-106}

\hangindent=1.5em\hangafter=1\noindent Gardner-Challis, Nelson, Bostock, J., Kozhevnikov, G., Sinclaire, M., Velja, J., Abate, A., and Griffin, C. ``When can we trust untrusted monitoring? A safety case sketch across collusion strategies.'' arXiv, September 21, 2026. \url{https://arxiv.org/abs/2602.20628}.\par

\phantomsection\label{source-ref-107}

\hangindent=1.5em\hangafter=1\noindent Geiping, Jonas, McLeish, S., Jain, N., Kirchenbauer, J., Singh, S., Bartoldson, B. R., Kailkhura, B., Bhatele, A., and Goldstein, T. ``Scaling up Test-Time Compute with Latent Reasoning: A Recurrent Depth Approach.'' arXiv, February 17, 2025. \url{https://arxiv.org/abs/2502.05171}.\par

\phantomsection\label{source-ref-108}

\hangindent=1.5em\hangafter=1\noindent Google DeepMind. ``Gemini 3 Pro Model Card.'' November 2025a, updated May 2026. \url{https://storage.googleapis.com/deepmind-media/Model-Cards/Gemini-3-Pro-Model-Card.pdf}.\par

\phantomsection\label{source-ref-109}

\hangindent=1.5em\hangafter=1\noindent Google DeepMind. ``Gemini 3 Pro Frontier Safety Framework Report.'' November 2025b. \url{https://storage.googleapis.com/deepmind-media/gemini/gemini_3_pro_fsf_report.pdf}.\par

\phantomsection\label{source-ref-110}

\hangindent=1.5em\hangafter=1\noindent Google DeepMind. ``Gemini 4 Argon Safety \& Security Report'', October 2026. \url{https://storage.googleapis.com/deepmind-media/gemini/gemini_4_safety_report.pdf}.\par

\phantomsection\label{source-ref-111}

\hangindent=1.5em\hangafter=1\noindent Greenblatt, Ryan, Shlegeris, B., Sachan, K., and Roger, F. ``AI Control: Improving Safety Despite Intentional Subversion.'' arXiv, updated July 23, 2024. \url{https://arxiv.org/abs/2312.06942}.\par

\phantomsection\label{source-ref-112}

\hangindent=1.5em\hangafter=1\noindent Greenblatt, Ryan, Westover, A., and Finnveden, L. ``Proposal for tracking the effects of architecture on monitorability.'' Redwood Research Blog, September 10, 2026. \url{https://blog.redwoodresearch.org/p/proposal-for-tracking-the-effects}. Accessed: 2026-10-08.\par

\phantomsection\label{source-ref-113}

\hangindent=1.5em\hangafter=1\noindent Guan, Melody Y., Wang, M., Carroll, M., Dou, Z., Wei, A. Y., Williams, M., Arnav, B., Huizinga, J., Kivlichan, I., Glaese, M., Pachocki, J., and Baker, B. ``Monitoring Monitorability.'' arXiv, December 20, 2025. \url{https://arxiv.org/abs/2512.18311}.\par

\phantomsection\label{source-ref-114}

\hangindent=1.5em\hangafter=1\noindent Hao, Shibo, Sukhbaatar, S., Su, D., Li, X., Hu, Z., Weston, J., and Tian, Y. ``Training Large Language Models to Reason in a Continuous Latent Space.'' August 23, 2026. \url{https://arxiv.org/abs/2412.06769}.\par

\phantomsection\label{source-ref-115}

\hangindent=1.5em\hangafter=1\noindent Kahn, Jeremy. ``Safety experts warn novel design of OpenAI's Astra model could make future AI agents harder to monitor.'' \emph{Fortune}, September 3, 2026. \url{https://fortune.com/2026/09/03/reports-openais-astra-model-uses-a-new-more-efficient-ai-architecture-alarms-ai-safety-experts-who-worry-the-method-makes-models-harder-to-control/}. Accessed: 2026-10-08.\par

\phantomsection\label{source-ref-116}

\hangindent=1.5em\hangafter=1\noindent Kinniment, Megan, Nix, S., Broadley, T., Wijk, H., and Parikh, N. ``Early work on monitorability evaluations''. METR. January 22, 2026. \url{https://metr.org/blog/2026-01-19-early-work-on-monitorability-evaluations/}.\par

\phantomsection\label{source-ref-117}

\hangindent=1.5em\hangafter=1\noindent Korbak, Tomek, Balesni, M., Barnes, E., Bengio, Y., Benton, J., Bloom, J., Chen, M., Cooney, A., Dafoe, A., Dragan, A., Emmons, S., Evans, O., Farhi, D., Greenblatt, R., Hendrycks, D., Hobbhahn, M., Hubinger, E., Irving, G., Jenner, E., Kokotajlo, D., Krakovna, V., Legg, S., Lindner, D., Luan, D., Madry, A., Michael, J., Nanda, N., Orr, D., Pachocki, J., Perez, E., Phuong, M., Roger, F., Saxe, J., Shlegeris, B., Soto, M., Steinberger, E., Wang, J., Zaremba, W., Baker, B., Shah, R., and Mikulik, V. ``Chain of Thought Monitorability: A New and Fragile Opportunity for AI Safety.'' arXiv, July 15, 2025, updated December 7, 2025. \url{https://arxiv.org/abs/2507.11473}.\par

\phantomsection\label{source-ref-118}

\hangindent=1.5em\hangafter=1\noindent Kutasov, Jonathan, Sun, Y., Colognese, P., van der Weij, T., Petrini, L., Zhang, C. B. C., Hughes, J., Deng, X., Sleight, H., Tracy, T., Shlegeris, B., and Benton, J. ``SHADE-Arena: Evaluating Sabotage and Monitoring in LLM Agents.'' arXiv, June 17, 2025, updated July 8, 2025. \url{https://arxiv.org/abs/2506.15740}.\par

\phantomsection\label{source-ref-119}

\hangindent=1.5em\hangafter=1\noindent Lanham, Tamera, Chen, A., Radhakrishnan, A., Steiner, B., Denison, C., Hernandez, D., Li, D., Durmus, E., Hubinger, E., Kernion, J., Lukošiūtė, K., Nguyen, K., Cheng, N., Joseph, N., Schiefer, N., Rausch, O., Larson, R., McCandlish, S., Kundu, S., Kadavath, S., Yang, S., Henighan, T., Maxwell, T., Telleen-Lawton, T., Hume, T., Hatfield-Dodds, Z., Kaplan, J., Brauner, J., Bowman, S. R., and Perez, E. ``Measuring faithfulness in chain-of-thought reasoning.'' arXiv, July 17, 2023. \url{https://arxiv.org/abs/2307.13702}.\par

\phantomsection\label{source-ref-120}

\hangindent=1.5em\hangafter=1\noindent Liu, Manqing, Williams-King, D., Caspary, I., Le, L., Whittingham, H., Radmard, P., Tice, C., and Young, E. J. ``Diagnosing Pathological Chain-of-Thought in Reasoning Models.'' arXiv, February 14, 2026, updated July 23, 2026. \url{https://arxiv.org/abs/2602.13904}.\par

\phantomsection\label{source-ref-121}

\hangindent=1.5em\hangafter=1\noindent Lu, Wenqan, Yang, Y., Lee, K., Li, Y., and Liu, E. ``Latent Chain-of-Thought? Decoding the Depth-Recurrent Transformer.'' In \emph{The First Workshop on the Application of LLM Explainability to Reasoning and Planning}, October 4, 2025. \url{https://openreview.net/forum?id=roIQdXMuEj}.\par

\phantomsection\label{source-ref-122}

\hangindent=1.5em\hangafter=1\noindent Mayne, Harry, Kang, J. S., Gould, D., Ramchandran, K., Mahdi, A., and Siegel, N. Y. ``A Positive Case for Faithfulness: LLM Self-Explanations Help Predict Model Behavior.'' arXiv, February 2, 2026, updated October 6, 2026. \url{https://arxiv.org/abs/2602.02639}.\par

\phantomsection\label{source-ref-123}

\hangindent=1.5em\hangafter=1\noindent Meek, Austin, Sprejer, E., Arcuschin, I., Brockmeier, A. J., and Basart, S. ``Measuring Chain-of-Thought Monitorability Through Faithfulness and Verbosity.'' arXiv, October 31, 2025, updated November 30, 2025. \url{https://arxiv.org/abs/2510.27378}.\par

\phantomsection\label{source-ref-124}

\hangindent=1.5em\hangafter=1\noindent Meinke, Alexander, Schoen, B., Scheurer, J., Balesni, M., Shah, R., and Hobbhahn, M. ``Frontier Models are Capable of In-context Scheming.'' arXiv, January 14, 2025. \url{https://arxiv.org/abs/2412.04984}.\par

\phantomsection\label{source-ref-125}

\hangindent=1.5em\hangafter=1\noindent METR. ``Frontier Risk Report (February to March 2026).'' May 19, 2026a. \url{https://metr.org/blog/2026-05-19-frontier-risk-report/}. Accessed: 2026-06-15.\par

\phantomsection\label{source-ref-126}

\hangindent=1.5em\hangafter=1\noindent METR. ``Brief independent investigation of agents' behavior, reasoning and collaboration in the OpenAI / Hugging Face hacking incident.'' August 26, 2026b. \url{https://metr.org/blog/2026-08-26-openai-hugging-face-incident-investigation/}. Accessed: 2026-10-08.\par

\phantomsection\label{source-ref-127}

\hangindent=1.5em\hangafter=1\noindent METR. ``Review of the `Misalignment in high-stakes settings' section in Anthropic's risk report (August 2026).'' October 20, 2026c. \url{https://metr.org/blog/2026-10-20-misalignment-risk-report-august-2026-review/}. Accessed: 2026-10-22.\par

\phantomsection\label{source-ref-128}

\hangindent=1.5em\hangafter=1\noindent METR. ``Red-Teaming OpenAI\textquotesingle s Internal Agent Monitoring Systems.'' October 22, 2026d. \url{https://metr.org/blog/2026-10-22-red-teaming-openai-agent-monitoring/}. Accessed: 2026-10-22.\par

\phantomsection\label{source-ref-129}

\hangindent=1.5em\hangafter=1\noindent OpenAI. ``Learning to reason with LLMs.'' September 12, 2024. \url{https://openai.com/index/learning-to-reason-with-llms/}. Accessed: 2026-04-24.\par

\phantomsection\label{source-ref-130}

\hangindent=1.5em\hangafter=1\noindent OpenAI. ``Detecting misbehavior in frontier reasoning models.'' March 10, 2025. \url{https://openai.com/index/chain-of-thought-monitoring/}. Accessed: 2026-10-08.\par

\phantomsection\label{source-ref-131}

\hangindent=1.5em\hangafter=1\noindent OpenAI. ``GPT-5.4 Thinking System Card.'' March 5, 2026a. \url{https://deploymentsafety.openai.com/gpt-5-4-thinking}.\par

\phantomsection\label{source-ref-132}

\hangindent=1.5em\hangafter=1\noindent OpenAI. ``GPT-5.6 System Card''. July 9, 2026b. \url{https://deploymentsafety.openai.com/gpt-5-6}.\par

\phantomsection\label{source-ref-133}

\hangindent=1.5em\hangafter=1\noindent OpenAI. ``The Hugging Face incident and the road ahead.'' August 26, 2026c. \url{https://openai.com/index/hugging-face-incident-and-the-road-ahead/}. Accessed: 2026-10-08.\par

\phantomsection\label{source-ref-134}

\hangindent=1.5em\hangafter=1\noindent OpenAI. ``GPT-6 Astra System Card.'' 3 September 2026d. \url{https://deploymentsafety.openai.com/gpt-6-astra}.\par

\phantomsection\label{source-ref-135}

\hangindent=1.5em\hangafter=1\noindent OpenAI. ``Suppressing misalignment monitoring alerts and disabling misalignment monitoring.'' October 20, 2026e. \url{https://alignment.openai.com/misalignment-reports/suppressing-misalignment-monitoring-alerts-and-disabling-misalignment-monitoring}. Accessed: 2026-10-21.\par

\phantomsection\label{source-ref-136}

\hangindent=1.5em\hangafter=1\noindent Pfau, Jacob, Merrill, W., and Bowman, S. R. ``Let's Think Dot by Dot: Hidden Computation in Transformer Language Models.'' arXiv, April 24, 2024. \url{https://arxiv.org/abs/2404.15758}.\par

\phantomsection\label{source-ref-137}

\hangindent=1.5em\hangafter=1\noindent Roger, Fabien and Greenblatt, R. ``Preventing Language Models From Hiding Their Reasoning.'' arXiv, October 31, 2023. \url{https://arxiv.org/abs/2310.18512}.\par

\phantomsection\label{source-ref-138}

\hangindent=1.5em\hangafter=1\noindent Schoen, Bronson, Nitishinskaya, E., Balesni, M., Højmark, A., Hofstätter, F., Scheurer, J., Meinke, A., Wolfe, J., van der Weij, T., Lloyd, A., Goldowsky-Dill, N., Fan, A., Matveiakin, A., Shah, R., Williams, M., Glaese, A., Barak, B., Zaremba, W., and Hobbhahn, M. ``Stress Testing Deliberative Alignment for Anti-Scheming Training.'' September 19, 2025. \url{https://arxiv.org/abs/2509.15541}.\par

\phantomsection\label{source-ref-139}

\hangindent=1.5em\hangafter=1\noindent Shah, Rohin and Dragan, A. ``The case for reasoning transparency.'' DeepMind Institute, September 16, 2026. \url{https://institute.deepmind.com/essays/the-case-for-reasoning-transparency/}. Accessed: 2026-10-08.\par

\phantomsection\label{source-ref-140}

\hangindent=1.5em\hangafter=1\noindent Skaf, Joey, Ibanez-Lissen, L., McCarthy, R., Watts, C., Georgiv, V., Whittingham, H., Gonzalez-Manzano, L., Lindner, D., Tice, C., Young, E. J., and Radmard, P. ``Large language models can learn and generalize steganographic chain-of-thought under process supervision.'' arXiv, December 4, 2025. URL \url{https://arxiv.org/abs/2506.01926}.\par

\phantomsection\label{source-ref-141}

\hangindent=1.5em\hangafter=1\noindent Tracy, Tyler, Potham, R., Kuhn, N., Heller, M., Khandelwal, A., Rushing, C., Lemoine, H., Brandao, M., Turlik, T., Hanson, A., Hills, J., Ngo, A., Rachum, R., Mitchell, N., Galperin, F., Sykes, O., Arnott, P., Prieto Lima, S., Giudice, C., Goldwater, M., Popp, D., de Wet, D., Castaing, R., Guo, Q., Marx, D., Shaffrey, B., Shenk, J., Milbradt, M., Meagher, H., Ahmed-Chowdhury, S., O'Connell, D., Canal, C., Shlegeris, B., and Bhatt, A. ``LinuxArena: A Control Setting for AI Agents in Live Production Software Environments.'' arXiv, April 27, 2026. \url{https://arxiv.org/abs/2604.15384}.\par

\phantomsection\label{source-ref-142}

\hangindent=1.5em\hangafter=1\noindent Turpin, Miles, Michael, J., Perez, E., and Bowman, S. R. ``Language Models Don\textquotesingle t Always Say What They Think: Unfaithful Explanations in Chain-of-Thought Prompting.'' December 9, 2023. \url{https://arxiv.org/abs/2305.04388}.\par

\phantomsection\label{source-ref-143}

\hangindent=1.5em\hangafter=1\noindent Van Hollen, Chris. Van Hollen presses OpenAI CEO Sam Altman on alarming new AI model claims, calls for risk assessment of AI capabilities. Press Release, September 2026. \url{https://www.vanhollen.senate.gov/news/press-releases/van-hollen-presses-openai-ceo-sam-altman-on-alarming-new-ai-model-claims-calls-for-risk-assessment-of-ai-capabilities}. Accessed: 2026-10-08. Archived November 1, 2026, at the Wayback Machine, \url{https://web.archive.org/web/20261101015352/https://www.vanhollen.senate.gov/news/press-releases/van-hollen-presses-openai-ceo-sam-altman-on-alarming-new-ai-model-claims-calls-for-risk-assessment-of-ai-capabilities}.\par

\phantomsection\label{source-ref-144}

\hangindent=1.5em\hangafter=1\noindent Wang, Xi, Cai, Z., Zhan, Z., Dong, H., Fan, Y., de Rosa, G., Pearce, T., and Langford, J. ``Full-bandwidth transformer.'' arXiv, October 1, 2026. \url{https://arxiv.org/abs/2608.08888}.\par

\phantomsection\label{source-ref-145}

\hangindent=1.5em\hangafter=1\noindent Yueh-Han, Chen, McCarthy, R., Lee, B. W., He, H., Kivlichan, I., Baker, B., Carroll, M., and Korbak, T. ``Reasoning Models Struggle to Control their Chains of Thought.'' arXiv, March 5, 2026. \url{https://arxiv.org/abs/2603.05706}.\par

\endgroup

\clearpage

\Needspace{6\baselineskip}

\section{\texorpdfstring{Appendix }{Appendix }}\label{appendix}

\Needspace{6\baselineskip}

\subsection{Sources Reviewed}\label{sources-reviewed}

The 21 candidate indicators were generated from a review of the following sources (among others): {\begingroup\hypersetup{linkcolor=MainTextLinkColor,urlcolor=MainTextLinkColor}\hyperref[source-ref-117]{Korbak et al. (2025)}\endgroup{}} on CoT monitorability as a fragile opportunity; {\begingroup\hypersetup{linkcolor=MainTextLinkColor,urlcolor=MainTextLinkColor}\hyperref[source-ref-97]{Baker et al. (2025)}\endgroup{}} on obfuscation under RL optimisation; {\begingroup\hypersetup{linkcolor=MainTextLinkColor,urlcolor=MainTextLinkColor}\hyperref[source-ref-113]{Guan et al. (2025)}\endgroup{}} on monitoring monitorability; {\begingroup\hypersetup{linkcolor=MainTextLinkColor,urlcolor=MainTextLinkColor}\hyperref[source-ref-102]{Delaney et al. (2025)}\endgroup{}} on policy options for preserving CoT monitorability; {\begingroup\hypersetup{linkcolor=MainTextLinkColor,urlcolor=MainTextLinkColor}\hyperref[source-ref-123]{Meek et al. (2025)}\endgroup{}} on measuring faithfulness and verbosity; {\begingroup\hypersetup{linkcolor=MainTextLinkColor,urlcolor=MainTextLinkColor}\hyperref[source-ref-120]{Liu et al. (2026)}\endgroup{}} on diagnosing pathological CoT; {\begingroup\hypersetup{linkcolor=MainTextLinkColor,urlcolor=MainTextLinkColor}\hyperref[source-ref-140]{Skaf et al. (2025)}\endgroup{}} on steganographic CoT under process supervision; {\begingroup\hypersetup{linkcolor=MainTextLinkColor,urlcolor=MainTextLinkColor}\hyperref[source-ref-122]{Mayne et al. (2026)}\endgroup{}} on faithfulness and behaviour prediction; {\begingroup\hypersetup{linkcolor=MainTextLinkColor,urlcolor=MainTextLinkColor}\hyperref[source-ref-95]{Anwar et al. (2025)}\endgroup{}} on information-theoretic analysis of CoT monitorability; {\begingroup\hypersetup{linkcolor=MainTextLinkColor,urlcolor=MainTextLinkColor}\hyperref[source-ref-138]{Schoen et al. (2025)}\endgroup{}} on stress testing deliberative alignment; {\begingroup\hypersetup{linkcolor=MainTextLinkColor,urlcolor=MainTextLinkColor}\hyperref[source-ref-103]{Emmons et al. (2025)}\endgroup{}} on when CoT helps; the OpenAI GPT-5.4 Thinking System Card ({\begingroup\hypersetup{linkcolor=MainTextLinkColor,urlcolor=MainTextLinkColor}\hyperref[source-ref-131]{OpenAI, 2026a}\endgroup{}}); the GPT-6 Astra System Card ({\begingroup\hypersetup{linkcolor=MainTextLinkColor,urlcolor=MainTextLinkColor}\hyperref[source-ref-134]{OpenAI, 2026d}\endgroup{}}); the Claude Fable 5.1 \& Claude Mythos 5.1 System Card ({\begingroup\hypersetup{linkcolor=MainTextLinkColor,urlcolor=MainTextLinkColor}\hyperref[source-ref-90]{Anthropic, 2026d}\endgroup{}}); the Gemini 4 Argon Safety \& Security Report ({\begingroup\hypersetup{linkcolor=MainTextLinkColor,urlcolor=MainTextLinkColor}\hyperref[source-ref-110]{Google DeepMind, 2026}\endgroup{}}); and METR's early monitorability evaluations ({\begingroup\hypersetup{linkcolor=MainTextLinkColor,urlcolor=MainTextLinkColor}\hyperref[source-ref-116]{Kinniment et al., 2026}\endgroup{}}).

\begin{landscape}\global\LandscapePagetrue

\Needspace{6\baselineskip}

\subsection{Full Scored Framework}\label{full-scored-framework}

All 21 candidate transparency measures, grouped by the monitorability condition each one bears on. I = Informativeness, T = Technology Readiness, EV = External Verifiability, DT = Developer Tractability, IF = Implementation Feasibility. Each is scored 1--5; \ensuremath{\Sigma} is the sum. Within each condition, the highest-scoring measure that is also feasible to report is selected as that condition's indicator (shaded), except for condition 1, which contains two indicators to understand whether the CoT is load-bearing for the model's complex reasoning capabilities, and condition 3, whose indicator is a suite of complementary probes. Where the top two tie on \ensuremath{\Sigma} (conditions 2 and 5), the tie is broken in favor of the more directly informative measure, as noted in {\begingroup\hypersetup{linkcolor=MainTextLinkColor,urlcolor=MainTextLinkColor}\hyperref[method-for-identifying-monitorability-indicators]{Section 3}\endgroup{}}.

\noindent\begin{minipage}{\linewidth}

\Needspace{6\baselineskip}

\subsubsection{Condition 1 --- The CoT Must Exist and be Necessary for Complex Reasoning}\label{condition-1-the-cot-must-exist-and-be-necessary-for-complex-reasoning}

\begingroup
\small
\setlength{\tabcolsep}{4pt}
\arrayrulecolor{TableGrid}
\setlength{\arrayrulewidth}{0.5pt}
\setlength{\tableusablewidth}{\dimexpr\linewidth-16\tabcolsep-9\arrayrulewidth\relax}
\setlength{\LTpost}{0pt}
\renewcommand{\arraystretch}{1.24}
\begin{tabular}{|>{\raggedright\arraybackslash}p{0.26500\tableusablewidth}|
>{\centering\arraybackslash}p{0.03100\tableusablewidth}|
>{\centering\arraybackslash}p{0.03100\tableusablewidth}|
>{\centering\arraybackslash}p{0.03100\tableusablewidth}|
>{\centering\arraybackslash}p{0.03100\tableusablewidth}|
>{\centering\arraybackslash}p{0.03100\tableusablewidth}|
>{\centering\arraybackslash}p{0.03100\tableusablewidth}|
>{\raggedright\arraybackslash}p{0.54900\tableusablewidth}|}
\hline
\rowcolor{TableHeading}
\textcolor{white}{\bfseries Measure} & \textcolor{white}{\bfseries I} & \textcolor{white}{\bfseries T} & \textcolor{white}{\bfseries EV} & \textcolor{white}{\bfseries DT} & \textcolor{white}{\bfseries IF} & \textcolor{white}{\bfseries \ensuremath{\Sigma}} & \textcolor{white}{\bfseries Justification} \\ \hline
\rowcolor{SelectedRow}
\textbf{Third-Party CoT Access} & 5 & 5 & 5 & 3 & 4 & \textbf{22} & \emph{Selected (indicator 1 / verification precondition).} Earliest external signal that text CoT still exists, and the access that lets third parties verify the other indicators. IP and misuse concerns bind tractability. \\ \hline
\rowcolor{SelectedRow}
\textbf{Base Architecture Specifications} & 4 & 4 & 3 & 4 & 2 & 17 & \emph{Selected (indicator 1 / verification precondition).} Determines whether text-based CoT exists and is necessary for the model's complex reasoning capabilities, but is sensitive product information. \\ \hline
\rowcolor{BodyRow}
Latent Reasoning Fraction & 5 & 3 & 2 & 3 & 3 & 16 & A rising fraction signals the transition to latent reasoning, but is not externally verifiable without CoT access. \\ \hline
\rowcolor{BodyRow}
Neuralese Translator Availability & 4 & 2 & 3 & 2 & 2 & 13 & Translation tools determine whether monitoring is possible with latent CoT; no validated translator currently exists. \\ \hline
\rowcolor{BodyRow}
Neuralese Detection Rate & 4 & 1 & 3 & 3 & 2 & 13 & Technology readiness = 1 is binding; no operational definition. \\ \hline
\rowcolor{BodyRow}
CoT Data Composition & 3 & 2 & 2 & 3 & 2 & 12 & Causal pathway to CoT quality not fully established. \\ \hline
\end{tabular}
\endgroup

\end{minipage}\par\medskip

\noindent\begin{minipage}{\linewidth}

\Needspace{6\baselineskip}

\subsubsection{Condition 2 --- The CoT Must Be in Readable Human Language}\label{condition-2-the-cot-must-be-in-readable-human-language}

\begingroup
\small
\setlength{\tabcolsep}{4pt}
\arrayrulecolor{TableGrid}
\setlength{\arrayrulewidth}{0.5pt}
\setlength{\tableusablewidth}{\dimexpr\linewidth-16\tabcolsep-9\arrayrulewidth\relax}
\setlength{\LTpost}{0pt}
\renewcommand{\arraystretch}{1.24}
\begin{tabular}{|>{\raggedright\arraybackslash}p{0.26500\tableusablewidth}|
>{\centering\arraybackslash}p{0.03100\tableusablewidth}|
>{\centering\arraybackslash}p{0.03100\tableusablewidth}|
>{\centering\arraybackslash}p{0.03100\tableusablewidth}|
>{\centering\arraybackslash}p{0.03100\tableusablewidth}|
>{\centering\arraybackslash}p{0.03100\tableusablewidth}|
>{\centering\arraybackslash}p{0.03100\tableusablewidth}|
>{\raggedright\arraybackslash}p{0.54900\tableusablewidth}|}
\hline
\rowcolor{TableHeading}
\textcolor{white}{\bfseries Measure} & \textcolor{white}{\bfseries I} & \textcolor{white}{\bfseries T} & \textcolor{white}{\bfseries EV} & \textcolor{white}{\bfseries DT} & \textcolor{white}{\bfseries IF} & \textcolor{white}{\bfseries \ensuremath{\Sigma}} & \textcolor{white}{\bfseries Justification} \\ \hline
\rowcolor{SelectedRow}
\textbf{Naturalness (Human Interpretability Scores)} & 5 & 3 & 4 & 3 & 3 & \textbf{18} & \emph{Selected (indicator 2).} Legibility of the reasoning affects human and AI monitors; not standardized. \\ \hline
\rowcolor{BodyRow}
CoT Coverage / Verbosity Trends & 3 & 3 & 4 & 4 & 4 & 18 & Whether the CoT externalizes all causal factors; framework exists but unadopted. A more indirect proxy for readability. \\ \hline
\end{tabular}
\endgroup

\end{minipage}\par\medskip

\clearpage

\noindent\begin{minipage}{\linewidth}

\Needspace{6\baselineskip}

\subsubsection{Condition 3 --- The CoT Must Be Faithful}\label{condition-3-the-cot-must-be-faithful}

\begingroup
\small
\setlength{\tabcolsep}{4pt}
\arrayrulecolor{TableGrid}
\setlength{\arrayrulewidth}{0.5pt}
\setlength{\tableusablewidth}{\dimexpr\linewidth-16\tabcolsep-9\arrayrulewidth\relax}
\setlength{\LTpost}{0pt}
\renewcommand{\arraystretch}{1.24}
\begin{tabular}{|>{\raggedright\arraybackslash}p{0.26500\tableusablewidth}|
>{\centering\arraybackslash}p{0.03100\tableusablewidth}|
>{\centering\arraybackslash}p{0.03100\tableusablewidth}|
>{\centering\arraybackslash}p{0.03100\tableusablewidth}|
>{\centering\arraybackslash}p{0.03100\tableusablewidth}|
>{\centering\arraybackslash}p{0.03100\tableusablewidth}|
>{\centering\arraybackslash}p{0.03100\tableusablewidth}|
>{\raggedright\arraybackslash}p{0.54900\tableusablewidth}|}
\hline
\rowcolor{TableHeading}
\textcolor{white}{\bfseries Measure} & \textcolor{white}{\bfseries I} & \textcolor{white}{\bfseries T} & \textcolor{white}{\bfseries EV} & \textcolor{white}{\bfseries DT} & \textcolor{white}{\bfseries IF} & \textcolor{white}{\bfseries \ensuremath{\Sigma}} & \textcolor{white}{\bfseries Justification} \\ \hline
\rowcolor{SelectedRow}
\textbf{Paraphrasing Tests (Steganography)} & 5 & 3 & 4 & 4 & 4 & \textbf{20} & \emph{Selected (as part of the suite for indicator 3).} Directly detects steganographic encoding by rewording the CoT and checking for a performance drop; established ({\begingroup\hypersetup{linkcolor=MainTextLinkColor,urlcolor=MainTextLinkColor}\hyperref[source-ref-119]{Lanham et al., 2023}\endgroup{}}) but not standardized. \\ \hline
\rowcolor{SelectedRow}
\textbf{Cue-Based Faithfulness Scores} & 4 & 3 & 4 & 4 & 4 & \textbf{19} & \emph{Selected (as part of the suite for indicator 3).} Probes post-hoc rationalization by checking whether the CoT acknowledges factors that influenced the answer; an established, Inspect-compatible method ({\begingroup\hypersetup{linkcolor=MainTextLinkColor,urlcolor=MainTextLinkColor}\hyperref[source-ref-142]{Turpin et al., 2023}\endgroup{}}), though not standardized. \\ \hline
\rowcolor{SelectedRow}
\textbf{Perturbation and Necessity Tests} & 4 & 3 & 4 & 4 & 4 & \textbf{19} & \emph{Selected (as part of the suite for indicator 3).} Probes scaffolding by intervening on the CoT to test whether it is load-bearing; demonstrated ({\begingroup\hypersetup{linkcolor=MainTextLinkColor,urlcolor=MainTextLinkColor}\hyperref[source-ref-123]{Meek et al., 2025}\endgroup{}}) but not standardized. \\ \hline
\rowcolor{BodyRow}
Post-Training Faithfulness Shifts & 5 & 3 & 3 & 3 & 4 & 18 & Change in faithfulness pre/post training; informative about training effects without revealing the procedures themselves. \\ \hline
\rowcolor{BodyRow}
Reinforcement Learning from Human Feedback/Reinforcement Learning from Artificial Intelligence Feedback on CoT & 4 & 2 & 2 & 3 & 3 & 14 & Whether feedback explicitly trains CoT quality; no standardised disclosure format. \\ \hline
\rowcolor{BodyRow}
Reinforcement Learning (RL) Environment Specifications & 4 & 2 & 2 & 3 & 2 & 13 & Specific RL setups that correlate with obfuscation; sensitive training information. \\ \hline
\rowcolor{BodyRow}
Optimization Pressure on CoT & 5 & 2 & 1 & 2 & 2 & 12 & Most empirically documented cause of obfuscation, but unverifiable externally and highly sensitive. \\ \hline
\rowcolor{BodyRow}
Monitorability-Focused Training & 4 & 2 & 1 & 2 & 2 & 11 & Unverifiable externally; sensitive training information. \\ \hline
\end{tabular}
\endgroup

\emph{The three shaded rows above jointly form the faithfulness suite (indicator 3).}

\end{minipage}\par\medskip

\clearpage

\noindent\begin{minipage}{\linewidth}

\Needspace{6\baselineskip}

\subsubsection{Condition 4 --- The Monitor Must Catch Harmful Behavior}\label{condition-4-the-monitor-must-catch-harmful-behavior}

\begingroup
\small
\setlength{\tabcolsep}{4pt}
\arrayrulecolor{TableGrid}
\setlength{\arrayrulewidth}{0.5pt}
\setlength{\tableusablewidth}{\dimexpr\linewidth-16\tabcolsep-9\arrayrulewidth\relax}
\setlength{\LTpost}{0pt}
\renewcommand{\arraystretch}{1.24}
\begin{tabular}{|>{\raggedright\arraybackslash}p{0.26500\tableusablewidth}|
>{\centering\arraybackslash}p{0.03100\tableusablewidth}|
>{\centering\arraybackslash}p{0.03100\tableusablewidth}|
>{\centering\arraybackslash}p{0.03100\tableusablewidth}|
>{\centering\arraybackslash}p{0.03100\tableusablewidth}|
>{\centering\arraybackslash}p{0.03100\tableusablewidth}|
>{\centering\arraybackslash}p{0.03100\tableusablewidth}|
>{\raggedright\arraybackslash}p{0.54900\tableusablewidth}|}
\hline
\rowcolor{TableHeading}
\textcolor{white}{\bfseries Measure} & \textcolor{white}{\bfseries I} & \textcolor{white}{\bfseries T} & \textcolor{white}{\bfseries EV} & \textcolor{white}{\bfseries DT} & \textcolor{white}{\bfseries IF} & \textcolor{white}{\bfseries \ensuremath{\Sigma}} & \textcolor{white}{\bfseries Justification} \\ \hline
\rowcolor{SelectedRow}
\textbf{Detection (Monitor Detection Rate)} & 5 & 3 & 3 & 4 & 4 & \textbf{19} & \emph{Selected (indicator 4).} Bottom-line lagging indicator; disclosures from multiple developers but no standardized benchmark. \\ \hline
\rowcolor{BodyRow}
Real-World Monitor Bypass Incidents & 5 & 2 & 3 & 4 & 4 & 18 & Most direct evidence of failure, with precedent from aviation/nuclear, but developers would push back on releasing it. \\ \hline
\rowcolor{BodyRow}
Internal Red-Team Findings & 5 & 3 & 2 & 3 & 3 & 16 & Directly tests monitor evasion, but internal by nature and carries a conflict of interest. \\ \hline
\end{tabular}
\endgroup

\end{minipage}\par\medskip

\noindent\begin{minipage}{\linewidth}

\Needspace{6\baselineskip}

\subsubsection{Condition 5 --- The CoT Must Add Value Over Action-Only Monitoring}\label{condition-5-the-cot-must-add-value-over-action-only-monitoring}

\begingroup
\small
\setlength{\tabcolsep}{4pt}
\arrayrulecolor{TableGrid}
\setlength{\arrayrulewidth}{0.5pt}
\setlength{\tableusablewidth}{\dimexpr\linewidth-16\tabcolsep-9\arrayrulewidth\relax}
\setlength{\LTpost}{0pt}
\renewcommand{\arraystretch}{1.24}
\begin{tabular}{|>{\raggedright\arraybackslash}p{0.26500\tableusablewidth}|
>{\centering\arraybackslash}p{0.03100\tableusablewidth}|
>{\centering\arraybackslash}p{0.03100\tableusablewidth}|
>{\centering\arraybackslash}p{0.03100\tableusablewidth}|
>{\centering\arraybackslash}p{0.03100\tableusablewidth}|
>{\centering\arraybackslash}p{0.03100\tableusablewidth}|
>{\centering\arraybackslash}p{0.03100\tableusablewidth}|
>{\raggedright\arraybackslash}p{0.54900\tableusablewidth}|}
\hline
\rowcolor{TableHeading}
\textcolor{white}{\bfseries Measure} & \textcolor{white}{\bfseries I} & \textcolor{white}{\bfseries T} & \textcolor{white}{\bfseries EV} & \textcolor{white}{\bfseries DT} & \textcolor{white}{\bfseries IF} & \textcolor{white}{\bfseries \ensuremath{\Sigma}} & \textcolor{white}{\bfseries Justification} \\ \hline
\rowcolor{SelectedRow}
\textbf{Uplift (CoT Monitoring Uplift, \ensuremath{\Delta}\textsubscript{1})} & 5 & 3 & 2 & 3 & 3 & \textbf{16} & \emph{Selected (indicator 5).} Quantifies the value of CoT access; disclosures from multiple developers demonstrate reportability, though it depends on CoT access and is not standardized. \\ \hline
\rowcolor{BodyRow}
Inference Compute Per Reasoning Token & 3 & 4 & 2 & 4 & 3 & 16 & Indirect indicator only; many factors affect compute. \\ \hline
\end{tabular}
\endgroup

\end{minipage}\par\medskip

\clearpage

\Needspace{6\baselineskip}

\subsection{Threat Models and the Conditions They Break}\label{threat-models-and-the-conditions-they-break}

Documented CoT failure modes from the literature, each mapped to the earliest condition in the chain that it breaks. This supports the exhaustiveness discussion in {\begingroup\hypersetup{linkcolor=MainTextLinkColor,urlcolor=MainTextLinkColor}\hyperref[a-framework-for-chain-of-thought-monitoring]{Section 2}\endgroup{}}: every threat we identified in the CoT literature breaks at least one of the five conditions. Each threat is placed at the earliest condition it breaks; some (e.g., steganography) bear on more than one.

\begingroup
\small
\setlength{\tabcolsep}{4pt}
\arrayrulecolor{TableGrid}
\setlength{\arrayrulewidth}{0.5pt}
\setlength{\tableusablewidth}{\dimexpr\linewidth-8\tabcolsep-5\arrayrulewidth\relax}
\setlength{\LTpost}{0pt}
\renewcommand{\arraystretch}{1.05}
\begin{longtable}{|>{\raggedright\arraybackslash}p{0.28000\tableusablewidth}|
>{\raggedright\arraybackslash}p{0.23000\tableusablewidth}|
>{\centering\arraybackslash}p{0.06500\tableusablewidth}|
>{\raggedright\arraybackslash}p{0.42500\tableusablewidth}|}
\hline
\rowcolor{TableHeading}
\textcolor{white}{\bfseries Threat Model} & \textcolor{white}{\bfseries Source(s)} & \textcolor{white}{\bfseries Cond.} & \textcolor{white}{\bfseries Mechanism} \\ \hline
\endfirsthead
\hline
\rowcolor{TableHeading}
\textcolor{white}{\bfseries Threat Model} & \textcolor{white}{\bfseries Source(s)} & \textcolor{white}{\bfseries Cond.} & \textcolor{white}{\bfseries Mechanism} \\ \hline
\endhead
\endfoot
\endlastfoot
\rowcolor{BodyRow}
Latent/Continuous-State Reasoning & {\begingroup\hypersetup{linkcolor=MainTextLinkColor,urlcolor=MainTextLinkColor}\hyperref[source-ref-121]{Lu et al., 2025}\endgroup{}}; {\begingroup\hypersetup{linkcolor=MainTextLinkColor,urlcolor=MainTextLinkColor}\hyperref[source-ref-107]{Geiping et al., 2025}\endgroup{}}; {\begingroup\hypersetup{linkcolor=MainTextLinkColor,urlcolor=MainTextLinkColor}\hyperref[source-ref-114]{Hao et al., 2026}\endgroup{}}; {\begingroup\hypersetup{linkcolor=MainTextLinkColor,urlcolor=MainTextLinkColor}\hyperref[source-ref-144]{Wang et al., 2026}\endgroup{}} & 1 & Reasoning leaves text entirely; no CoT to inspect; CoT is inspectable but might not be the only reasoning mechanism used by the model. \\ \hline
\rowcolor{BodyRow}
Architectural Move Away From Text CoT (Efficiency Pressure) & {\begingroup\hypersetup{linkcolor=MainTextLinkColor,urlcolor=MainTextLinkColor}\hyperref[source-ref-102]{Delaney et al., 2025}\endgroup{}} & 1 & Removes the language bottleneck; CoT ceases to exist. \\ \hline
\rowcolor{BodyRow}
Language Drift to Non-Standard English & {\begingroup\hypersetup{linkcolor=MainTextLinkColor,urlcolor=MainTextLinkColor}\hyperref[source-ref-138]{Schoen et al., 2025}\endgroup{}} & 2 & CoT exists but is no longer readable. \\ \hline
\rowcolor{BodyRow}
Neuralese/Model-Convenient Compression & {\begingroup\hypersetup{linkcolor=MainTextLinkColor,urlcolor=MainTextLinkColor}\hyperref[source-ref-138]{Schoen et al., 2025}\endgroup{}} & 2 & Token use unnatural to humans; semantics not extractable. \\ \hline
\rowcolor{BodyRow}
Post-Hoc Rationalization & {\begingroup\hypersetup{linkcolor=MainTextLinkColor,urlcolor=MainTextLinkColor}\hyperref[source-ref-142]{Turpin et al., 2023}\endgroup{}} & 3 & Reasoning reconstructed from the answer; not load-bearing. \\ \hline
\rowcolor{BodyRow}
CoT-as-Scaffolding (Forward-Pass Compute \ensuremath{\ne} Emitted Tokens) & {\begingroup\hypersetup{linkcolor=MainTextLinkColor,urlcolor=MainTextLinkColor}\hyperref[source-ref-136]{Pfau et al., 2024}\endgroup{}} & 3 & Text is a prop for internal computation, not a window. \\ \hline
\rowcolor{BodyRow}
Steganographic/Encoded Reasoning & {\begingroup\hypersetup{linkcolor=MainTextLinkColor,urlcolor=MainTextLinkColor}\hyperref[source-ref-137]{Roger \& Greenblatt, 2023}\endgroup{}}; {\begingroup\hypersetup{linkcolor=MainTextLinkColor,urlcolor=MainTextLinkColor}\hyperref[source-ref-140]{Skaf et al., 2025}\endgroup{}} & 3 & Hidden content in word choice / token order; passes naive reading. \\ \hline
\rowcolor{BodyRow}
Obfuscation Under Reinforcement Learning Optimization Against a Monitor & {\begingroup\hypersetup{linkcolor=MainTextLinkColor,urlcolor=MainTextLinkColor}\hyperref[source-ref-97]{Baker et al., 2025}\endgroup{}} & 3 & Optimization pressure trains the CoT to evade the monitor. \\ \hline
\rowcolor{BodyRow}
Pathological CoT (Necessity/Paraphrasability/Substantivity Failures) & {\begingroup\hypersetup{linkcolor=MainTextLinkColor,urlcolor=MainTextLinkColor}\hyperref[source-ref-120]{Liu et al., 2026}\endgroup{}} & 3 & Diagnostic suite shows the CoT is not doing the computational work. \\ \hline
\rowcolor{BodyRow}
Weak/Low-Capability Monitor & {\begingroup\hypersetup{linkcolor=MainTextLinkColor,urlcolor=MainTextLinkColor}\hyperref[source-ref-97]{Baker et al., 2025}\endgroup{}}; {\begingroup\hypersetup{linkcolor=MainTextLinkColor,urlcolor=MainTextLinkColor}\hyperref[source-ref-113]{Guan et al., 2025}\endgroup{}} & 4 & Faithful CoT exists, but the monitor fails to flag harm. \\ \hline
\rowcolor{BodyRow}
Monitor--Model Collusion & {\begingroup\hypersetup{linkcolor=MainTextLinkColor,urlcolor=MainTextLinkColor}\hyperref[source-ref-111]{Greenblatt et al., 2024}\endgroup{}}; {\begingroup\hypersetup{linkcolor=MainTextLinkColor,urlcolor=MainTextLinkColor}\hyperref[source-ref-106]{Gardner-Challis et al., 2026}\endgroup{}} & 4 & Untrusted monitor colludes; detection compromised. \\ \hline
\rowcolor{BodyRow}
Monitor Is Not Robust to CoT Rationalization/Biased Reasoning & {\begingroup\hypersetup{linkcolor=MainTextLinkColor,urlcolor=MainTextLinkColor}\hyperref[source-ref-91]{Anthropic, 2026e}\endgroup{}} & 4 & Monitor fails to flag harm because it is convinced by the model's CoT that its activities are permissible. \\ \hline
\rowcolor{BodyRow}
In-Context Scheming Evading Monitoring & {\begingroup\hypersetup{linkcolor=MainTextLinkColor,urlcolor=MainTextLinkColor}\hyperref[source-ref-124]{Meinke et al., 2025}\endgroup{}} & 4 & Model reasons about evading the specific monitor. \\ \hline
\rowcolor{BodyRow}
Shrinking CoT-vs-Action Gap (Action-Only Catches Up) & {\begingroup\hypersetup{linkcolor=MainTextLinkColor,urlcolor=MainTextLinkColor}\hyperref[source-ref-96]{Arnav et al., 2025}\endgroup{}}; {\begingroup\hypersetup{linkcolor=MainTextLinkColor,urlcolor=MainTextLinkColor}\hyperref[source-ref-113]{Guan et al., 2025}\endgroup{}} & 5 & CoT access stops adding marginal monitoring value. \\ \hline
\rowcolor{BodyRow}
CoT Quality Too Poor to Help Over Actions & {\begingroup\hypersetup{linkcolor=MainTextLinkColor,urlcolor=MainTextLinkColor}\hyperref[source-ref-113]{Guan et al., 2025}\endgroup{}} & 5 & Even a capable monitor gains nothing from reading the CoT. \\ \hline
\end{longtable}
\endgroup

\end{landscape}\global\LandscapePagefalse

\end{document}